\documentclass[journal]{IEEEtran} 

\usepackage{xr-hyper} 

\usepackage{cite}

\ifCLASSINFOpdf
  \usepackage[pdftex]{graphicx}
  \usepackage{epstopdf} 
\else
\fi
\usepackage{amsmath}
\usepackage[ruled,vlined,linesnumbered]{algorithm2e}

\DontPrintSemicolon
\SetAlgoNoEnd
\SetKwInput{KwInput}{Input}
\SetKwInput{KwOutput}{Output}
  \usepackage[caption=false,font=footnotesize]{subfig}

\usepackage{array}
\usepackage{tabularx}

\usepackage{url}

\usepackage{amsthm}

\newtheorem{theorem}{Theorem}
\newtheorem{corollary}{Corollary}[theorem]
\newtheorem{lemma}[theorem]{Lemma}

\newtheorem{definition}{Definition}

\newtheoremstyle{myremark}
{}
{}
{}
{0pt}
{\bfseries}
{.}
{ }
{\thmname{#1}\thmnumber{ #2}: \thmnote{#3}}
\theoremstyle{myremark}

\newtheoremstyle{remarkshort}
{}
{}
{}
{0pt}
{\bfseries}
{.}
{ }
{\thmname{#1}\thmnumber{ #2}}

\theoremstyle{remarkshort}

\theoremstyle{remarkshort}

\usepackage{amssymb,amsmath}
\usepackage{xspace}
\usepackage{bm}
\usepackage{bbm}
\usepackage{mathtools}

\let\originalleft\left
\let\originalright\right
\renewcommand{\left}{\mathopen{}\mathclose\bgroup\originalleft}
\renewcommand{\right}{\aftergroup\egroup\originalright}

\renewcommand{\vec}[1]{{\mathbf{#1}}} 

\newcommand{\eqdef}{\mathrel{\mathop:}=}

\DeclareMathOperator*{\argmax}{\mathrm{argmax}}

\newcommand{\snr}{{\mathsf{SNR}}}

\newcommand{\inr}{{\mathsf{INR}}}

\def\vc{{\vec{c}}}

\def\vp{{\vec{p}}}

\def\vs{{\vec{s}}}

\def\vlambda{{\bm{\lambda}}}

\newcommand{\IDLE}{\mathrm{ID}}
\newcommand{\mnmax}{m_n^{\mathrm{max}}}
\usepackage{xspace}
\usepackage[acronym,nogroupskip,nonumberlist,nopostdot]{glossaries}
\makeglossaries

\usepackage{xcolor}

\newacronym{snr}{SNR}{signal-to-noise ratio}
\newacronym{sinr}{SINR}{signal-to-interference-plus-noise ratio}
\newacronym{inr}{INR}{interference-to-noise ratio}
\newacronym{sir}{SIR}{signal-to-interference ratio}
\newacronym{sqr}{SQR}{signal-to-quantization-noise ratio}
\newacronym{sqnr}{SQNR}{signal-to-quantization-plus-noise ratio}
\newacronym{ian}{IAN}{interference as noise}
\newacronym{ber}{BER}{bit error rate}
\newacronym{pn}{PN}{pseudorandom noise}
\newacronym{bfsk}{BFSK}{binary frequency shift keying}
\newacronym{fh}{FH}{frequency-hopped}
\newacronym{fh-bfsk}{FH-BFSK}{frequency-hopped binary frequency shift keying}
\newacronym{crc}{CRC}{cyclic redundancy check}
\newacronym{isi}{ISI}{intersymbol interference}
\newacronym{dsss}{DSSS}{direct-sequence spread spectrum}
\newacronym{ofdm}{OFDM}{orthogonal frequency-division multiplexing}
\newacronym{ofdma}{OFDMA}{orthogonal frequency-division multiple access}
\newacronym{sdr}{SDR}{software-defined radio}
\newacronym{tx}{TX}{transmitter}
\newacronym{rx}{RX}{receiver}
\newacronym{fdd}{FDD}{frequency-division duplexing}
\newacronym{tdd}{TDD}{time-division duplexing}
\newacronym{fdma}{FDMA}{frequency-division multiple access}
\newacronym{tdma}{TDMA}{time-division multiple access}
\newacronym{sdma}{SDMA}{space-division multiple access}
\newacronym[plural=MPCs]{mpc}{MPC}{multipath component}
\newacronym{mui}{MUI}{multi-user interference}
\newacronym{lsb}{LSB}{least significant bit}
\newacronym{jcas}{JCAS}{joint communication and sensing}

\newacronym{qam}{QAM}{quadrature amplitude modulation}
\newacronym{mqam}{MQAM}{M-ary quadrature amplitude modulation}

\newacronym{ls}{LS}{least-squares}
\newacronym{lms}{LMS}{least mean squares}
\newacronym{rls}{RLS}{recursive least-squares}
\newacronym{rzf}{RZF}{regularized zero-forcing}
\newacronym{mmse}{MMSE}{minimum mean square error}
\newacronym{lmmse}{LMMSE}{linear minimum mean square error}
\newacronym{mse}{MSE}{mean square error}
\newacronym{fft}{FFT}{fast Fourier transform}
\newacronym{dft}{DFT}{discrete Fourier transform}
\newacronym{dtft}{DTFT}{discrete-time Fourier transform}
\newacronym{ctft}{CTFT}{continuous-time Fourier transform}
\newacronym{ml}{ML}{machine learning}
\newacronym[plural=NNs]{nn}{NN}{neural network}
\newacronym[plural=RNNs]{rnn}{RNN}{recurrent neural network}
\newacronym[plural=ADCs]{adc}{ADC}{analog-to-digital converter}
\newacronym[plural=DACs]{dac}{DAC}{digital-to-analog converter}
\newacronym[plural=FPGAs]{fpga}{FPGA}{field-programmable gate array}
\newacronym{evm}{EVM}{error vector magnitude}
\newacronym{enob}{ENOB}{effective number of bits}
\newacronym{zf}{ZF}{zero-forcing}
\newacronym{rv}{r.v.}{random variable}
\newacronym{omp}{OMP}{orthogonal matching pursuit}
\newacronym{svd}{SVD}{singular value decomposition}

\newacronym{sdp}{SDP}{semidefinite programming}
\newacronym{psd}{PSD}{positive semidefinite}
\newacronym{nsd}{NSD}{negative semidefinite}

\newacronym{ks}{K-S}{Kolmogorov-Smirnov}

\newacronym{mad}{MAD}{median absolute deviation around the median}

\newacronym{agc}{AGC}{automatic gain control}
\newacronym{rf}{RF}{radio frequency}
\newacronym{if}{IF}{intermediate frequency}
\newacronym{los}{LOS}{line-of-sight}
\newacronym{nlos}{NLOS}{non-line-of-sight}
\newacronym{ple}{PLE}{path loss exponent}
\newacronym[plural=dB,firstplural=decibels (dB)]{db}{dB}{decibel}
\newacronym[plural=dBm,firstplural=decibel milliwatts (dBm)]{dbm}{dBm}{decibel milliwatts}
\newacronym{pa}{PA}{power amplifier}
\newacronym{lna}{LNA}{low noise amplifier}
\newacronym{vga}{VGA}{variable gain amplifier}
\newacronym{cw}{CW}{continuous wave}
\newacronym{papr}{PAPR}{peak-to-average power ratio}
\newacronym{usrp}{USRP}{Universal Software Radio Peripheral}
\newacronym{irr}{IRR}{image rejection ratio}
\newacronym{lo}{LO}{local oscillator}
\newacronym{vm}{VM}{vector modulator}
\newacronym{mmwave}{mmWave}{millimeter-wave}
\newacronym{eirp}{EIRP}{effective isotropic radiated power}
\newacronym{rsrp}{RSRP}{reference signal received power}

\newacronym{csma}{CSMA}{carrier-sense multiple access}
\newacronym{csmaca}{CSMA/CA}{carrier-sense multiple access with collision avoidance}
\newacronym{csmacd}{CSMA/CD}{carrier-sense multiple access with collision detection}
\newacronym{mac}{MAC}{medium access control}
\newacronym{phy}{PHY}{physical layer}
\newacronym{4g}{4G}{fourth generation}
\newacronym{lte}{LTE}{Long-Term Evolution}
\newacronym{4glte}{4G LTE}{\gls{4g} \gls{lte}}
\newacronym{5g}{5G}{fifth generation}
\newacronym{nr}{NR}{New Radio}
\newacronym{5gnr}{5G NR}{5G New Radio}
\newacronym{ieee}{IEEE}{Institute of Electrical and Electronics Engineers}
\newacronym{wifi}{Wi-Fi}{IEEE 802.11}
\newacronym{lan}{LAN}{local area network}
\newacronym{wlan}{WLAN}{wireless local area network}
\newacronym[plural=BSs]{bs}{BS}{base station}
\newacronym[plural=SBSs]{sbs}{SBS}{small-cell base station}
\newacronym[plural=FD-SBSs]{fdsbs}{FD-SBS}{\gls{fd}-enabled \gls{sbs}}
\newacronym[plural=MBSs]{mbs}{MBS}{macrocell base station}
\newacronym[plural=UEs]{ue}{UE}{user equipment}
\newacronym{ul}{UL}{uplink}
\newacronym{dl}{DL}{downlink}
\newacronym{qos}{QoS}{Quality of Service}
\newacronym{fcc}{FCC}{Federal Communications Commission}
\newacronym{iab}{IAB}{integrated access and backhaul}
\newacronym{fab}{FAB}{fixed access and backhaul}
\newacronym{hetnet}{HetNet}{heterogeneous network}

\newacronym{siso}{SISO}{single-input single-output}
\newacronym{mimo}{MIMO}{multiple-input multiple-output}
\newacronym{sumimo}{SU-MIMO}{single-user \gls{mimo}}
\newacronym{mumimo}{MU-MIMO}{multi-user \gls{mimo}}
\newacronym{bf}{BF}{beamforming}
\newacronym{ca}{CA}{constant amplitude}
\newacronym{ula}{ULA}{uniform linear array}
\newacronym{upa}{UPA}{uniform planar array}
\newacronym[\glslongpluralkey={angles of arrival}]{aoa}{AoA}{angle of arrival}
\newacronym[\glslongpluralkey={angles of departure}]{aod}{AoD}{angle of departure}
\newacronym{dof}{DoF}{degrees of freedom}
\newacronym{csi}{CSI}{channel state information}
\newacronym{csit}{CSIT}{\gls{csi} at the transmitter}
\newacronym{csir}{CSIR}{\gls{csi} at the receiver}
\newacronym{cs}{CS}{compressed sensing}

\newacronym{fd}{FD}{in-band full-duplex}
\newacronym{hd}{HD}{half-duplex}
\newacronym{si}{SI}{self-interference}
\newacronym{sic}{SIC}{self-interference cancellation}
\newacronym{soi}{SoI}{signal of interest}
\newacronym{asic}{A-SIC}{analog \acrlong{sic}}
\newacronym{dsic}{D-SIC}{digital \gls{sic}}
\newacronym{star}{STAR}{simultaneous transmit and receive}
\newacronym{warp}{WARP}{Wireless Open-Access Research Platform}
\newacronym{bfc}{BFC}{beamforming cancellation}
\newacronym{ipi}{IPI}{inter-panel-interference}
\newacronym{ipic}{IPIC}{inter-panel-interference cancellation}

\newacronym{qcqp}{QCQP}{quadratically-constrained quadratic programming}
\newacronym{pdf}{PDF}{probability density function}
\newacronym{cdf}{CDF}{cumulative density function}
\newacronym{iid}{i.i.d.}{independently and identically distributed}

\newacronym{elf}{ELF}{extremely low frequency}
\newacronym{slf}{SLF}{super low frequency}
\newacronym{ulf}{ULF}{ultra low frequency}
\newacronym{vlf}{VLF}{very low frequency}
\newacronym{lf}{LF}{low frequency}
\newacronym{mf}{MF}{medium frequency}
\newacronym{hf}{HF}{high frequency}
\newacronym{vhf}{VHF}{very high frequency}
\newacronym{uhf}{UHF}{ultra high frequency}
\newacronym{shf}{SHF}{super high frequency}
\newacronym{ehf}{EHF}{extremely high frequency}
\newacronym{thf}{THF}{tremendously high frequency}

\newacronym{wncg}{WNCG}{Wireless Networking and Communications Group}
\newacronym{linc}{LINC}{Laboratory of Informatics, Networks, and Communications}
\newacronym{ut}{UT Austin}{The University of Texas at Austin}
\newacronym{uiuc}{UIUC}{University of Illinois at Urbana-Champaign}
\newacronym{usc}{USC}{University of Southern California}
\newacronym{mit}{MIT}{Massachusetts Institute of Technology}
\newacronym{berkeley}{UC Berkeley}{University of California, Berkeley}
\newacronym{osu}{OSU}{Ohio State University}

\newcommand{\mmwave}{\gls{mmwave}\xspace}

\newtheorem{proposition}{Proposition}

\newtheorem{assumption}{Assumption}

\newcommand{\figref}[1]{\figurename~\ref{#1}}

\usepackage{mathtools}
\usepackage{comment}

\usepackage{enumitem}

\usepackage{pgfplots}
\pgfplotsset{compat=newest}
\usetikzlibrary{plotmarks}
\usetikzlibrary{arrows.meta}
\usepgfplotslibrary{patchplots}
\usepackage{grffile}
\pgfplotsset{plot coordinates/math parser=false}

\definecolor{uclablue}{RGB}{39,116,174}
\definecolor{uclabluedarkest}{RGB}{0,59,92}
\definecolor{uclabluedarker}{RGB}{0,85,135}
\definecolor{uclabluelighter}{RGB}{139,184,232}
\definecolor{uclabluelightest}{RGB}{218,235,254}

\definecolor{uclagold}{RGB}{255,209,0}
\definecolor{uclagolddarker}{RGB}{255,199,44}
\definecolor{uclagolddarkest}{RGB}{255,184,28}

\definecolor{uclaviolet}{RGB}{130,55,165}
\definecolor{uclamagenta}{RGB}{255,0,165}

\begin{document}

%
\title{Scheduling Algorithms for Millimeter-Wave Full-Duplex Integrated Access and Backhaul}

%
%
%

\author{%
Sungho Cho, Changhee Joo, and Ian P.~Roberts%
\thanks{Sungho Cho and Ian P.~Roberts are with the Department of Electrical and Computer Engineering,
University of California, Los Angeles, CA 90095 USA,
Email: \{shcho1304, ianroberts\}@ucla.edu.}%
\thanks{Changhee Joo is with the Department of Computer Science and Engineering, Korea University, Seoul, 02841, South Korea.
Email: changhee@korea.ac.kr.}%
}

\maketitle

\begin{abstract}
Integrated access and backhaul (IAB) is a cost-effective way to deploy millimeter-wave cellular networks, but under half-duplex (HD) operation, it exhibits poor latency and throughput scaling, as an IAB node's panels are typically scheduled to either all transmit or all receive in a given time slot to avoid self-interference. 
Upgrading IAB nodes with full-duplex capability (FD) allows their panels to be scheduled more flexibly, but doing so optimally across an entire network is challenging since self-interference couples link rates with scheduling decisions across nodes.
In this work, we first develop a distributed dynamic programming algorithm that resolves this coupling by passing messages up and down the IAB tree, computing throughput-optimal back-pressure schedules in near-linear time. 
To reduce the control overhead associated with this approach, we also propose Floor-based Reduced-Overhead MaxWeight (FROM), which instead allows each node to schedule its panels independent of how its child nodes are scheduled. 
FROM accomplishes this by computing a floor on outgoing backhaul rates and scheduling each node using only local queue and channel states and its parent node's decision. 
We characterize the downlink stability region of this class of reduced-overhead policies and show that it spans at least half of the global stability region. 
We further prove that FROM stabilizes a guaranteed fraction of this region and identify conditions under which it achieves throughput optimality. 
Numerical results show that full-duplex IAB networks support higher traffic loads and lower delay than conventional half-duplex ones, and that FROM approaches the performance of global scheduling, even outperforming it when accounting for control overhead.
\end{abstract}


\glsresetall

\section{Introduction} \label{sec:introduction}
The demand for high data rates and increased network capacity has motivated the use of \mmwave spectrum in fifth-generation (5G) cellular networks.
However, mmWave links are susceptible to blockage and propagation losses, making reliable coverage challenging \cite{xue2024survey}.
These limitations motivate dense deployments of base stations that shorten transmission distances and provide additional connectivity options.
Such deployments, in turn, require high-capacity backhaul connections, but installing fiber at every base station can be costly and impractical \cite{tezergil2022wireless,wen2024shaping}.

To reduce the cost of providing fiber backhaul to densely deployed base stations, 3GPP introduced integrated access and backhaul (IAB), which allows base stations to relay traffic over wireless links instead of requiring a fiber connection at every site.
Each IAB node provides access to user equipment (UE) and connects to an IAB donor through one or more wireless backhaul hops, enabling broad, reliable coverage \cite{wen2024shaping}.
Unlike fiber backhaul, wireless backhaul is susceptible to interference from other transmitting panels, requiring careful coordination of access and backhaul transmissions \cite{polese2020integrated,sadovaya2022integrated}.

IAB nodes can employ multiple antenna panels, each consisting of an antenna array and covering a different angular sector \cite{promponas2025optimization}.
Beamforming allows each panel to direct transmissions toward its receivers in the sector and substantially reduce interference to UEs in other sectors \cite{tezergil2022wireless}.
However, signals from a node's transmitting panels can still interfere with its receiving panels during simultaneous transmission and reception, causing self-interference (SI) \cite{roberts2023spatial}.
Half-duplex (HD) operation avoids this self-interference by separating transmission and reception in time, whereas full-duplex (FD) operation permits both simultaneously.

In networks with such interference characteristics, two factors can affect link rates: channel conditions and the set of concurrently active links.
Although these factors can change link rates, several studies formulate scheduling using fixed rates for links.
\cite{joo2009understanding} analyzed greedy maximal scheduling, which repeatedly selects the remaining link with the largest queue and excludes conflicting links, and proved that it achieves the full capacity region on tree networks when the rates are constant.
Using a Jackson-network model, \cite{gupta2023system} optimized network utility under throughput and probabilistic latency constraints under the fixed-rate model.
Under an interference-free FD model, \cite{yuan2020optimal} represented feasible simultaneous schedules through graph matchings and jointly optimized traffic routing and the fractions of time assigned to those transmissions.
For Gaussian 1-2-1 relay networks, \cite{ezzeldin2021gaussian} derived the information-theoretic approximate capacity of multi-hop FD relay networks and polynomial-time algorithms for computing the corresponding schedules.
For wireless broadcast networks, including an FD setting without self-interference, \cite{kose2021novel} exploited polynomial-time maximum-weight scheduling on claw-free conflict graphs.
For multi-hop mmWave networks with MU-MIMO radios, \cite{gomezcuba2020optimal} formulated backpressure scheduling as a maximum-weight directed bipartite subgraph problem under fixed power allocation and negligible inter-link interference.

Channel states can vary over time due to fading and blockage, which can weaken the received signal or temporarily interrupt a link.
Several studies consider these variations by adapting scheduling decisions to the current link rates.
To be robust to blockages, \cite{dogan2025reinforcement} combined proactive path selection with reinforcement learning that adjusts traffic allocation across the selected paths.
Considering RF-chain constraints and time-varying link rates, \cite{gopalam2022distributed} developed a distributed algorithm based on dynamic programming that computes maximum-weight schedules with time-varying channels.
Under HD operation, \cite{pagin2022resource} developed a semi-centralized resource allocation framework that computes weights from channel and buffer information and selects schedules through maximum-weight matching on a spanning tree.
Using neighboring links' queue lengths and current rates, \cite{joo2016distributed} developed distributed greedy scheduling that gives priority to larger link weights while avoiding conflicts.

In the aforementioned studies, the current rates are treated as given when the scheduler selects the active links, and interference is modeled through constraints that prevent conflicting links from being activated together rather than through schedule-dependent changes in link rates.
In contrast, other studies model how interference from other active links changes the rate of a link and account for this dependence when selecting simultaneous transmissions.
\cite{choi2014distributed} accounted for scheduling-dependent rates by representing each physical link as multiple virtual links, each with a fixed rate and a corresponding SINR requirement.
Their distributed scheduler uses a three-way handshake to ensure that all active links satisfy these requirements under cumulative interference and achieves throughput optimality.
\cite{sadovaya2024delay} formulated delay-aware joint routing and link scheduling as a Markov decision process and developed a genetic algorithm that accounts for interference among simultaneous transmissions.
For self-backhauled small cells, \cite{goyal2017scheduling} jointly selected HD or FD transmissions and adjusted transmit powers to balance the utility gained from concurrent transmissions against the interference they cause.
For FD-enabled mmWave IAB networks with D2D communication, \cite{ma2024coalition} changed which links share the same frequency resources by considering both a link's utility gain and the interference-induced utility changes of other affected links and optimized the utilities using a game-theoretic approach.
For FD self-backhauled vehicular networks, \cite{fang2022resource} jointly considered user association, frequency-resource assignment, and transmit powers under reliability and latency constraints.
The framework alternated between resource assignment and power allocation to account for their coupled effects on interference and performance.

Scheduling in a multi-panel FD IAB network must balance the flexibility from relaxing the restriction of simultaneous transmission and reception and the reduction in rate caused by self-interference.
In this FD IAB environment, however, the aforementioned works do not establish throughput-optimal scheduling that jointly accounts for time-varying link rates and schedule-dependent rates.
We develop a distributed dynamic programming algorithm that resolves these coupled decisions across the IAB tree and computes maximum-weight schedules while accounting for self-interference by passing messages throughout the tree.

IAB networks can be organized as a spanning tree or a directed acyclic graph (DAG) \cite{polese2020integrated}.
A DAG allows an IAB node to maintain multiple parents for alternative forwarding paths, whereas a spanning tree assigns a single parent to each non-donor node.
Although a parent-link failure may interrupt packet delivery in a spanning tree, the topology can be reconfigured when an alternative parent is available \cite{pagin2022resource}.
The network can therefore operate on one spanning tree and switch to another after a failure, which motivates our focus on scheduling access and backhaul transmissions over a spanning-tree topology.

However, the network-wide message propagation can still raise control overhead.
Thus, to reduce the overhead, several studies proposed reduced-overhead scheduling policies using only local information \cite{gupta2009low, gopalam2022distributed, gopalam2023distributed}.
\cite{gupta2009low} used each link's own and neighboring queue lengths to determine a random waiting time before attempting transmission, deferring transmission if an interfering link transmitted first.
For HD IAB networks, \cite{gopalam2022distributed} characterized the stability region achievable by local policies and proposed a local max-weight policy.
Alternatively, \cite{gopalam2023distributed} used locally available traffic and rate information to determine reservation lengths and neighboring reservation information to select nonconflicting future slots.

The time-varying channel state and rate dependence introduce an additional difficulty under such local information constraints.
A parent may need to select an outgoing backhaul link before knowing the receiving child's simultaneous transmissions, even though those transmissions determine the link rate.
Thus, the same interference structure that can be exploited for distributed optimization also creates uncertainty when scheduling decisions use only local information.
Our objective is to establish a reduced overhead scheduling policy utilizing this structure and to characterize the stability of the policy.
We therefore develop Floor-based Reduced-Overhead MaxWeight (FROM), which uses each node's queues, channel states, and neighbor information.
FROM addresses the uncertainty of the decision of children using the minimum rate supported over the child's permitted transmission choices.
We establish a stability guarantee for this policy and examine how its reduced information requirements affect the achievable stability region.

The main contributions of this paper are summarized as follows.
\begin{enumerate}
    \item We derive a distributed dynamic programming algorithm that maximizes the weighted sum of link rates while accounting for schedule-dependent self-interference.
    The algorithm exchanges and reuses subtree scheduling results to reduce computational complexity.
    \item We characterize the stability region achievable by a class of reduced overhead scheduling policies and bound its gap from the global stability region.
    This analysis quantifies the performance limitations imposed by the local information constraints.
    \item We propose FROM and prove that it stabilizes a fraction of the reduced overhead policies' stability region.
    \item Numerical results demonstrate that FD operation supports a wider stability region than HD operation.
    They also show that the gap between the empirical stability regions of the proposed reduced overhead scheduling policy and the global scheduling policies is small and the reduced overhead policy is robust to control delay.
\end{enumerate}

The remainder of this paper is organized as follows.
Section \ref{sec:system-model} introduces the IAB network model, self-interference model, queueing model, and scheduling constraints.
Section \ref{sec:scheduling} develops the distributed dynamic programming algorithm for maximum-weight scheduling and presents the global back-pressure policy and its stability properties.
Section \ref{sec:reduced-overhead-mw} characterizes the stability region of reduced overhead policies, introduces FROM, and establishes its downlink stability guarantee.
Section \ref{sec:numerical-results} compares the stability and delay performance of the scheduling policies through numerical experiments.
Finally, Section \ref{sec:conclusion} concludes the paper.
Detailed derivations and proofs are provided in the supplementary materials.

\section{System Model}\label{sec:system-model}

\begin{figure}[!t]
    \centering
    \subfloat[IAB network topology.]{%
        \includegraphics[width=\linewidth]{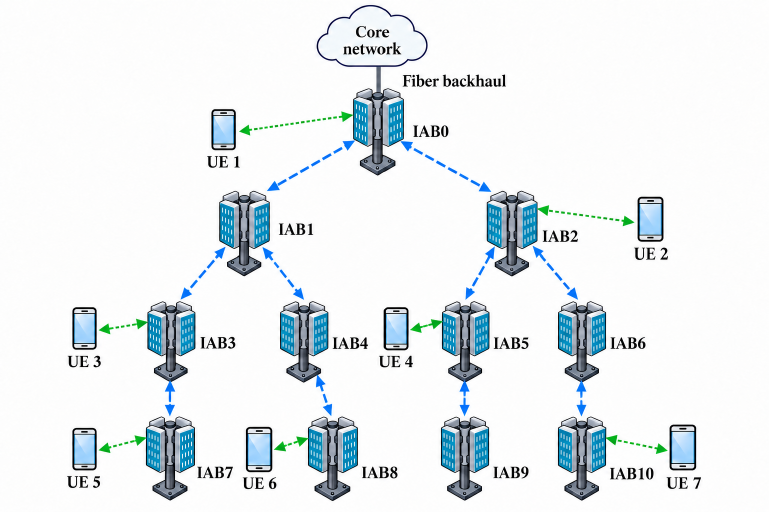}%
        \label{fig:iab-topology}%
    }
    \par\vspace{0.5em}
    \subfloat[Structure of IAB.]{%
        \includegraphics[width=0.4\linewidth]{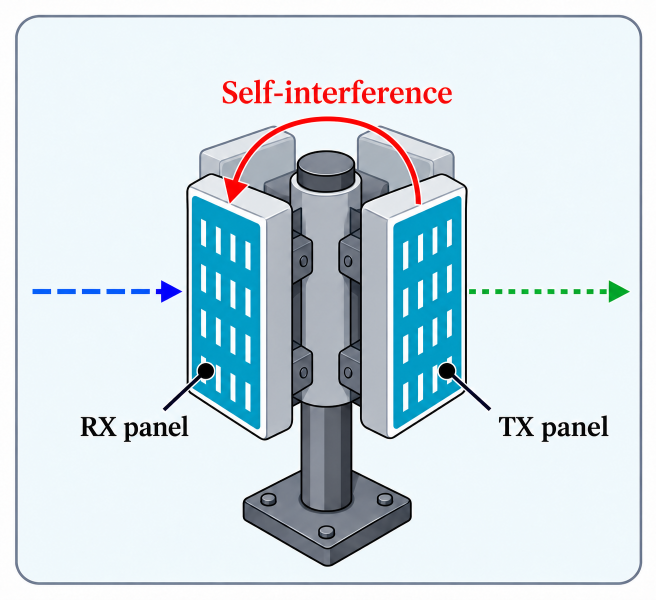}%
        \label{fig:iab-si}%
    }
    \caption{Multi-panel full-duplex IAB network.}
    \label{fig:iab-system}
\end{figure}

This work considers an FD IAB network with a tree topology, denoted by $\mathcal{G} = (\mathcal{B} \cup \mathcal{U}, \mathcal{L}, r)$, where $\mathcal{B}$ is the set of IAB nodes (gNBs) with $|\mathcal{B}| = B$, $\mathcal{U}$ is the disjoint set of user equipment (UE) with $|\mathcal{U}| = U$, and $r \in \mathcal{B}$ is the donor gNB.
The set of links, $\mathcal{L}$, contains the directed uplink and downlink communication links associated with its edges.
For every non-root IAB $n \in \mathcal{B} \setminus \{r\}$, $\mathrm{pr}(n) \in \mathcal{B}$ denotes its unique parent.
For every IAB $n \in \mathcal{B}$, $\mathrm{ch}(n) \subseteq \mathcal{B} \cup \mathcal{U}$ denotes its set of children.
Each IAB $n$ carries a set of panels $\mathcal{P}_n \eqdef \{P_n^1, P_n^2, \ldots, P_n^{p_n}\}$, and $\mathcal{P} \eqdef \bigcup_{n \in \mathcal{B}} \mathcal{P}_n$ is the set of all panels.
The map $I: \mathcal{P} \rightarrow \mathcal{B}$ returns the IAB hosting a given panel.
For each backhaul, the same panel is used in both directions at each endpoint.
We assume that each UE lies in the serving area $\mathcal{A}_{P_n^j}$ of exactly one panel $P_n^j$ thanks to the directivity of mmWave.
We additionally assume that each transmitting panel contributes the same SI to each receiving panel at that IAB so that the aggregate SI is proportional to the number of transmitting panels.

A link is an ordered pair $\ell : A(\ell) \rightarrow D(\ell)$, where $A(\ell), D(\ell) \in \mathcal{U} \cup \mathcal{P}$ are its transmitting and receiving endpoints, respectively.
The link set is partitioned as $\mathcal{L} = \mathcal{L}_{\mathrm{UL}} \cup \mathcal{L}_{\mathrm{DL}}$ and $\mathcal{L} = \mathcal{L}_{\mathrm{AC}} \cup \mathcal{L}_{\mathrm{BH}}$.
Uplinks point toward the root, whereas downlinks point away from it.
An access link in $\mathcal{L}_{\mathrm{AC}}$ connects a UE and an IAB, and a backhaul link in $\mathcal{L}_{\mathrm{BH}}$ connects two IABs.
For a non-root IAB $n$, its parent backhaul consists of the downlink $b_n^{\mathrm{DL}}$ from $\mathrm{pr}(n)$ to $n$ and the uplink $b_n^{\mathrm{UL}}$ from $n$ to $\mathrm{pr}(n)$; we write $b_n$ for this pair.
Both directions use panel $P_n^1$ at node $n$.

Let $\mathcal{L}_n$ denote the set of links incident on the panels of IAB $n$.
Its transmitting and receiving link sets are $\mathcal{L}_n^{\mathrm{TX}} \eqdef \{\ell \in \mathcal{L}_n : A(\ell) \in \mathcal{P}_n\}$ and $\mathcal{L}_n^{\mathrm{RX}} \eqdef \{\ell \in \mathcal{L}_n : D(\ell) \in \mathcal{P}_n\}$, respectively.
For a panel $k \in \mathcal{P}_n$, define $\mathcal{L}_{n,k} \eqdef \{\ell \in \mathcal{L}_n : A(\ell)=k \text{ or } D(\ell)=k\}$, $\mathcal{L}_{n,k}^{\mathrm{TX}} \eqdef \{\ell \in \mathcal{L}_{n,k} : A(\ell)=k\}$, and $\mathcal{L}_{n,k}^{\mathrm{RX}} \eqdef \{\ell \in \mathcal{L}_{n,k} : D(\ell)=k\}$.

\subsection{Structure of IAB}

We use a slotted time model with slots indexed by $t \in \mathbb{Z}_+$.
In each slot, the scheduler selects a binary activation $s_\ell(t) \in \{0,1\}$ for each $\ell \in \mathcal{L}$, where $s_\ell(t)=1$ indicates that link $\ell$ is scheduled.
The schedule is denoted by $\vs(t) \eqdef (s_\ell(t))_{\ell \in \mathcal{L}}$.
Each panel can support at most one active link per slot.
Consequently, the number of transmitting panels at IAB $n$ is
\begin{align}\label{eq:tx-panel-count}
    m_n(\vs,t) \eqdef \sum_{\ell \in \mathcal{L}_n^{\mathrm{TX}}} s_\ell(t).
\end{align}

For a link received at an IAB, let $n_\ell \eqdef I(D(\ell))$ denote its receiving IAB.
An access downlink is received by a UE and is free from SI.
Every other link is received at an IAB and is affected by the transmitting panels at that IAB.
The transmit-panel count seen by link $\ell$ is therefore
\begin{align}\label{eq:link-tx-count}
    m_\ell(\vs,t) \eqdef \begin{cases}
        0, & \ell \in \mathcal{L}_{\mathrm{AC}} \cap \mathcal{L}_{\mathrm{DL}},\\
        m_{n_\ell}(\vs,t), & \text{otherwise}.
    \end{cases}
\end{align}

For access links, channel conditions vary over time due to fading.
For backhaul links between stationary IAB nodes, we assume that channel conditions remain constant across non-outage slots.
Outages caused by tracking errors, beam misalignment, or blockage make a link unavailable for scheduling.
The channel process is assumed to be stationary and ergodic.

We assume that modulation and coding schemes (MCSs) with a finite set of supported spectral efficiencies are used.
The calculated spectral efficiency is rounded down to the largest supported MCS spectral efficiency that does not exceed it.
For each $m$, we first project the SINR-based rate onto a supported MCS level and then obtain the corresponding SNR value by inverting the Shannon rate expression using the same $m$ and INR.
We denote this value by $\snr_\ell(t)$ for access links and by $\snr_\ell$ for non-outage backhaul links, with its dependence on $m$ and INR omitted for notational simplicity.
For each fixed $m$, $\snr_\ell$ remains constant across non-outage slots under the backhaul channel assumption above.
If no MCS level is supported, the corresponding SNR value is set to zero.
Let $\mathcal M$ denote the finite set of SNR states over all links and candidate transmit-panel counts.

For non-outage access and backhaul links, the service rates, measured in bits, are respectively
\begin{align}
\label{eq:capacity}
\begin{aligned}
R_\ell(m,t)
&= \mathrm{BW}\,T_s \log_2\!\bigl(1+\Gamma_\ell(m,t)\bigr),\\
R_\ell^{\mathrm{BH}}(m)
&= \mathrm{BW}\,T_s \log_2\!\left(1+\frac{\snr_\ell}{1+m\,\inr_{n_\ell}}\right),
\end{aligned}
\end{align}
where $\mathrm{BW}$ is the bandwidth, $T_s$ is the slot duration, and the access-link SINR is
\begin{align}
\label{eq:sinr}
\Gamma_\ell(m,t)
=
\begin{cases}
\snr_\ell(t), & \ell\in\mathcal L_{\mathrm{AC}}\cap\mathcal L_{\mathrm{DL}},\\[4pt]
\dfrac{\snr_\ell(t)}{1+m\,\inr_{n_\ell}}, & \ell\in\mathcal L_{\mathrm{AC}}\cap\mathcal L_{\mathrm{UL}}.
\end{cases}
\end{align}
The term $\inr_n$ is the interference-to-noise ratio incurred by one transmitting panel to a receiving panel at IAB $n$.
Based on the statistical properties of SI, we assume that $\inr_n$ remains constant across slots \cite{roberts_lonestar}.
Thus, $R_\ell^{\mathrm{BH}}(m)$ is constant across non-outage slots for each fixed $m$.

When using a common rate notation for all links, we write $R_\ell(m,t)=R_\ell^{\mathrm{BH}}(m)$ for a non-outage backhaul link.
Under schedule $\vs(t)$, the offered service on link $\ell$ is $R_\ell(m_\ell(\vs,t),t)s_\ell(t)$, with $s_\ell(t)=0$ for links in outage.

Let $\gamma_{\min}$ denote the minimum SINR supported by the MCS.
For a backhaul link $\ell\in\mathcal L_n^{\mathrm{RX}}\cap\mathcal L_{\mathrm{BH}}$, the maximum transmit-panel count supporting reception is
\begin{align}
\label{eq:link-tx-limit}
m_\ell^{\max}\eqdef\max\left\{m\in\{0,\ldots,p_{n_\ell}-1\}:\frac{\snr_\ell}{1+m\,\inr_{n_\ell}}\ge\gamma_{\min}\right\}.
\end{align}
The upper bound $p_{n_\ell}-1$ reserves one panel for reception.

To support backhaul reception irrespective of which incoming link is selected, we use a common transmit-panel limit that satisfies the minimum SINR requirement of every incoming backhaul link in its non-outage state.
Thus, for an IAB $n$ with at least one such reception link, define
\begin{align}
\label{eq:node-tx-limit}
\mnmax
\eqdef
\min_{\ell\in\mathcal L_n^{\mathrm{RX}}\cap\mathcal L_{\mathrm{BH}}}
m_\ell^{\max}.
\end{align}

This constraint is only applied when IAB $n$ schedules at least one reception to allow transmitting when the interference is sufficiently high and the IAB does not receive any packets, i.e., if $\sum_{\ell \in \mathcal{L}_n^{\mathrm{RX}}} s_\ell(t)>0$, $m_n(\vs, t) \leq \mnmax$.

Each panel has a single RF chain and can be active on at most one link per slot.
For $n\in\mathcal{B},\ k\in\mathcal{P}_n, t\in\mathbb{Z}_+$,
\begin{align}\label{eq:panel-constraint}
    \sum_{\ell \in \mathcal{L}_{n,k}} s_\ell(t)\leq 1.
\end{align}

\subsection{Queueing Model}

Traffic is organized into flows.
A flow $f$ follows the unique ordered path from its source $\mathrm{src}(f)$ to its destination $\mathrm{dst}(f)$.
The flow set is partitioned as $\mathcal{F}=\mathcal{F}_{\mathrm{UL}}\cup\mathcal{F}_{\mathrm{DL}}$.
An uplink flow originates at a UE and terminates at the donor, whereas a downlink flow originates at the donor and terminates at a UE.
We write $n\in f$ and $\ell\in f$ when node $n$ and link $\ell$, respectively, lie on the path of $f$.
The set of flows traversing link $\ell$ is $\mathcal{F}_\ell\eqdef\{f\in\mathcal{F}:\ell\in f\}$.
Each node $n\in f\setminus\{\mathrm{dst}(f)\}$ has a separate queue of $q_n^f(t)$ bits for flow $f$ at the beginning of slot $t$.
Data leave the network upon reaching their destination, and we set $q_n^f(t)=0$ whenever $n\notin f$ or $n=\mathrm{dst}(f)$.
External arrivals occur only at the source of each flow.
For each flow $f \in \mathcal{F}$, the exogenous arrival process $\{\lambda_f(t)\}^\infty_{t=0}$ is an ergodic and stationary process, as assumed in \cite{neely2003dynamic, georgiadis2006resource}.
Let $d_n^f(t)\in\mathbb{Z}_+$ denote the actual number of bits removed from the queue of flow $f$ at node $n$, and let $a_n^f(t)\in\mathbb{Z}_+$ denote the number added to that queue during slot $t$.
Actual departures satisfy $0\leq d_n^f(t)\leq q_n^f(t)$, and the total departures over each link cannot exceed its offered service.
For $n\in f\setminus\{\mathrm{dst}(f)\}$, the queueing equation is defined as
\begin{align}\label{eq:queue-evolution}
    q_n^f(t+1) = \bigl(q_n^f(t)-d_n^f(t)\bigr)^+ + a_n^f(t),
\end{align}
where $(x)^+\eqdef\max\{x,0\}$.

For a downlink flow $f\in\mathcal{F}_{\mathrm{DL}}$,
\begin{align}\label{eq:dl-arrivals}
    a_n^f(t) = \begin{cases}
        \lambda_f(t), & n=r,\\
        d_{\mathrm{pr}(n)}^f(t), & n\in f\cap(\mathcal{B}\setminus\{r\}),\\
        0, & \text{otherwise}.
    \end{cases}
\end{align}
For an uplink flow $f\in\mathcal{F}_{\mathrm{UL}}$,
\begin{align}\label{eq:ul-arrivals}
    a_n^f(t) = \begin{cases}
        \lambda_f(t), & n=\mathrm{src}(f),\\
        \displaystyle\sum_{c\in\mathrm{ch}(n)\cap f}d_c^f(t), & n\in f\cap(\mathcal{B}\setminus\{r\}),\\
        0, & \text{otherwise}.
    \end{cases}
\end{align}

\subsection{Scheduling Policy}

We will first define the vector notation.
Let
\[
    \boldsymbol Q(t) \eqdef \bigl(q_n^f(t)\bigr)_{n\in\mathcal{B}\cup\mathcal{U},\,f\in\mathcal{F}}
\]
denote the queue-length vector, and let
\[
    \snr(t) \eqdef \bigl(\snr_\ell(t)\bigr)_{\ell\in\mathcal{L}}
\]
denote the link-state vector, taking values in $\mathcal{M}$.
Let $\mathcal{S}(\snr(t))$ denote the set of binary schedules satisfying the aforementioned constraints, such as \eqref{eq:panel-constraint}, with $s_\ell(t)=0$ for every link with $R_\ell(0,t)=0$.
A scheduling policy should choose a schedule $\vs(t)$ from this feasible set, $\mathcal{S}$.
In this work, we consider the following two classes of stationary scheduling policies.

\begin{definition}\label{def:stationary-policy}
A stationary scheduling policy selects $\vs(t)\in\mathcal{S}(\snr(t))$ only based on the current state $(\boldsymbol Q(t),\snr(t))$.
\begin{itemize}
    \item A deterministic stationary policy specifies a fixed mapping from
    \[
        (\boldsymbol Q,\snr) \in \mathbb{Z}_+^{|\mathcal{F}|(|\mathcal{B}|+|\mathcal{U}|)} \times\mathcal{M}
    \]
    to a feasible schedule $\vs\in\mathcal{S}(\snr)$.
    \item A randomized stationary policy assigns to each state $(\boldsymbol Q, \snr)$ a fixed probability distribution $P_{\boldsymbol Q,\snr}$ over $\mathcal{S}(\snr)$ and draws
    \[
        \vs(t)\sim P_{\boldsymbol Q(t),\snr(t)}.
    \]
\end{itemize}
\end{definition}

\section{Dynamic Programming for Maximum Weighted Schedule} \label{sec:scheduling}

We consider global scheduling policies that select a feasible schedule using queue and channel information from across the network.
We first develop a distributed algorithm for maximum-weight scheduling, then characterize the global stability region and establish that the resulting back-pressure policy stabilizes every arrival rate inside this region.

\subsection{Finding the Maximum Weighted Schedule} \label{sec:mws}

The maximum-weight scheduling problem is
\begin{align}
\label{eq:mws}
\vs^\star \eqdef \argmax_{\vs \in \mathcal S}
\sum_{\ell \in \mathcal L}
w_\ell\bigl(m_\ell(\vs,t),t\bigr)s_\ell,
\end{align}
where the weights depend on the link rates and, depending on the scheduling objective, the queue lengths.
For example, using link rates as weights gives the sum-rate objective \cite{ma2024coalition}, whereas combining rates with differential queue backlogs gives the back-pressure objective \cite{gopalam2022distributed}.

In FD operation, the rate of a link received at an IAB depends on that IAB's transmit-panel count, so activating one link can change the weights of other links.
Directly optimizing these coupled decisions can require evaluating many scheduling combinations.
We therefore use dynamic programming to reuse optimal subtree values and avoid repeatedly evaluating the same combinations.
Let $\mathcal G_n$ denote the induced subgraph containing $n$ and all its descendants, and let $\mathcal C_n \eqdef \{c \in \mathcal B : \mathrm{pr}(c)=n\}$ denote its IAB children.
For brevity, links whose weights are non-positive are automatically eliminated from $\mathcal{L}$.
In an upward pass, each node combines the optimal values reported by its children with its local link weights to compute and report the optimal values for $\mathcal G_n$.
These reports capture the contributions of descendant subtrees without requiring the parent to enumerate their schedules.
A subsequent downward pass selects the corresponding link activations.

For a node $n$, let $\mathcal{L}(\mathcal{G}_n)$ denote the set of links of the subgraph $\mathcal{G}_n$.
Since the weighted sum over $\mathcal{G}_n$ depends on the schedule of its parent backhaul $b_n$, we represent this backhaul state by the mode $\theta_n$, which is
\begin{equation}
\label{eq:bn-mode}
\theta_n=
\begin{cases}
T, & s_{b_n^{\mathrm{DL}}}=1,\\
R, & s_{b_n^{\mathrm{UL}}}=1,\\
\IDLE, & s_{b_n^{\mathrm{DL}}}=s_{b_n^{\mathrm{UL}}}=0.
\end{cases}
\end{equation}
For feasible modes, the optimal weight of $\mathcal G_n$, including its parent backhaul, is
\begin{equation}
\label{eq:dp-nodeval}
\begin{aligned}
\nu_n^T
&=\max_{0\leq m\leq\mnmax}
\bigl[w_{b_n^{\mathrm{DL}}}(m)+\Phi_n^T(m)\bigr],\\
\nu_n^R(m')
&=w_{b_n^{\mathrm{UL}}}(m')
+\max_{1\leq m\leq p_n}\Phi_n^R(m),\\
\nu_n^{\IDLE}
&=\max_{0\leq m\leq p_n}\Phi_n^{\IDLE}(m).
\end{aligned}
\end{equation}
Here, $\Phi_n^\theta(m)$ is the maximum weight achievable in $\mathcal G_n$ over feasible subtree schedules with inherited mode $\theta$ and total TX-panel count $m$ at node $n$:
\begin{equation}
\Phi_n^\theta(m)
=
\max_{\substack{
\vs\in\mathcal S\\
\theta_n=\theta,\;m_n=m
}}
\sum_{\ell\in\mathcal L(\mathcal G_n)}
w_\ell\bigl(m_\ell(\vs)\bigr)s_\ell.
\end{equation}
In mode $R$, one of the $m$ TX panels is used for the parent backhaul, leaving $m-1$ TX panels for links within $\mathcal G_n$ and the parent backhaul scheduling is fixed by $\theta$.

The optimal value of \eqref{eq:mws} is $\nu_r^\IDLE$, since the root has no parent backhaul.
The corresponding schedules of each $\nu_n^{\theta_n}$ are denoted as $\vs_n^T, \vs_n^R(m'), \vs_n^\IDLE$.
We will compute the three $\nu_n^T, \nu_n^R(m'), \nu_n^\IDLE$ at a leaf node and then extend the computation to an intermediate node.

\subsubsection{Leaf Node} \label{sec:dp-leaf}

Let $n$ be a leaf, i.e., an IAB whose child links are all access links.
By \eqref{eq:panel-constraint}, each panel activates at most one link.
A transmitting panel selects the TX link with the largest weight, whereas a non-transmitting panel selects the RX link with the largest weight or remains idle if all candidate RX weights are not positive.
The maximum weights achievable by panel $k$ in these two cases, denoted by $\psi_{n,k}^T$ and $\psi_{n,k}^R(m)$, respectively, are given by the following expressions, where $m$ is the total number of transmitting panels at $n$:
\begin{align}
\psi_{n,k}^T
&=\max_{\ell\in\mathcal L_{n,k}^{\mathrm{TX}}
\cap\mathcal L_{\mathrm{AC}}}
w_\ell,
\label{eq:leaf-psiT}\\
\psi_{n,k}^R(m)
&=\bigl[
\max_{\ell\in\mathcal L_{n,k}^{\mathrm{RX}}
\cap\mathcal L_{\mathrm{AC}}}
w_\ell(m)
\bigr]^+.
\label{eq:leaf-psiR}
\end{align}
For $\mnmax<m\leq p_n$, reception is not permitted, so $\psi_{n,k}^R(m)=0$.
\begin{lemma}
\label{lem:leaf-dp}
At a leaf $n$, the three values of \eqref{eq:dp-nodeval} are
\begin{align}
    \nu_n^T
    &=\max_{0\leq m\leq\mnmax}
    \Bigg[w_{b_n^{\mathrm{DL}}}(m)+\sum_{k\in\mathcal{P}_n\setminus\{P_n^1\}}\psi_{n,k}^R(m)\nonumber\\
    &\qquad+\max_{\substack{P\subseteq\mathcal{P}_n\setminus\{P_n^1\}\\|P|=m}}
    \sum_{k\in P}\big\{\psi_{n,k}^T-\psi_{n,k}^R(m)\big\}\Bigg],
    \label{eq:leaf-T}\\
    \nu_n^R(m')
    &=w_{b_n^{\mathrm{UL}}}(m')+\max_{1\leq m\leq p_n}
    \Bigg[\sum_{k\in\mathcal{P}_n\setminus\{P_n^1\}}\psi_{n,k}^R(m)\nonumber\\
    &\qquad+\max_{\substack{P\subseteq\mathcal{P}_n\setminus\{P_n^1\}\\|P|=m-1}}
    \sum_{k\in P}\big\{\psi_{n,k}^T-\psi_{n,k}^R(m)\big\}\Bigg],
    \label{eq:leaf-R}\\
    \nu_n^{\IDLE}
    &=\max_{0\leq m\leq p_n}\Bigg[\sum_{k\in\mathcal{P}_n}\psi_{n,k}^R(m) +\max_{\substack{P\subseteq\mathcal{P}_n\\|P|=m}}
    \sum_{k\in P}\big\{\psi_{n,k}^T-\psi_{n,k}^R(m)\big\}\Bigg].
    \label{eq:leaf-phi}
\end{align}
\end{lemma}
\begin{proof}
    See Appendix \ref{app:dp-derivation} of the supplementary material.    
\end{proof}

\subsubsection{Intermediate Node} \label{sec:dp-inter}

Let $n$ be an intermediate node with at least one child IAB, so that its child backhaul links are given by $\{b_c\}_{c\in\mathcal{C}_n}$.
Equations \eqref{eq:leaf-T}--\eqref{eq:leaf-phi} are reused with the definitions of $\psi_{n,k}^T$ and $\psi_{n,k}^R(m)$ below and the idle contribution of children whose backhauls share the reserved panel $P_n^1$.
$\psi_{n,k}^T$ and $\psi_{n,k}^R(m)$ are given in the following lemma.
\begin{lemma}
Let $B_{n,k}$ denote the total optimal weight of the child subtrees associated with panel $k$ when their connecting backhauls are idle:
\begin{equation}
B_{n,k}\eqdef
\sum_{\substack{c\in\mathcal C_n\\
b_c^{\mathrm{DL}}\in\mathcal L_{n,k}}}
\nu_c^{\IDLE}.
\end{equation}
The panel values are
\begin{align}
\psi_{n,k}^T
&=B_{n,k}
+\max_{\ell\in\mathcal L_{n,k}^{\mathrm{TX}}
\cap\mathcal L(\mathcal G_n)}
(\tau_\ell-\phi_\ell),
\label{eq:inter-T}\\
\psi_{n,k}^R(m)
&=B_{n,k}
+\bigl[
\max_{\ell\in\mathcal L_{n,k}^{\mathrm{RX}}
\cap\mathcal L(\mathcal G_n)}
(\rho_\ell(m)-\phi_\ell)
\bigr]^+.
\label{eq:inter-R}
\end{align}
The expression for $\psi_{n,k}^R(m)$ applies to $0\leq m\leq\mnmax$.
For $\mnmax<m\leq p_n$, reception is not permitted, so $\psi_{n,k}^R(m)=B_{n,k}$.
The link values are defined by
\begin{align}
\label{eq:inter-linkval}
\tau_\ell
&=\begin{cases}
w_\ell,
&\ell\in\mathcal L_{\mathrm{AC}},\\
\nu_c^T,
&\ell=b_c^{\mathrm{DL}},\ c\in\mathcal C_n,
\end{cases}\\
\rho_\ell(m)
&=\begin{cases}
w_\ell(m),
&\ell\in\mathcal L_{\mathrm{AC}},\\
\nu_c^R(m),
&\ell=b_c^{\mathrm{UL}},\ c\in\mathcal C_n,
\end{cases}\\
\phi_\ell
&=\begin{cases}
0,
&\ell\in\mathcal L_{\mathrm{AC}},\\
\nu_c^{\IDLE},
&\ell\in\{b_c^{\mathrm{DL}},b_c^{\mathrm{UL}}\},\
c\in\mathcal C_n.
\end{cases}
\end{align}
\end{lemma}
\begin{proof}
    See Appendix \ref{app:dp-derivation} of the supplementary material.
\end{proof}

For $\theta\in\{T,R\}$, panel $P_n^1$ is occupied by the backhaul between $n$ and its parent.
Since each panel can activate at most one link, any child backhaul sharing this panel must remain inactive.
Each affected child $c$ can still schedule transmissions within its subtree, but with its parent backhaul inactive, its inherited mode is $\IDLE$.
The optimal weight contributed by that subtree is therefore $\nu_c^{\IDLE}$.
\begin{align}
    \Phi_n^\theta(m)
    &=\sum_{\substack{c\in\mathcal{C}_n\\b_c^{\mathrm{DL}}\in\mathcal{L}_{n,P_n^1}}}\nu_c^{\IDLE}
    +\sum_{k\in\mathcal{P}_n\setminus\{P_n^1\}}\psi_{n,k}^R(m)\nonumber\\
    &\quad+\max_{\substack{P\subseteq\mathcal{P}_n\setminus\{P_n^1\}\\|P|=m-\mathbf{1}\{\theta=R\}}}
    \sum_{k\in P}\big\{\psi_{n,k}^T-\psi_{n,k}^R(m)\big\}.\label{eq:inter-phi}
\end{align}
For $\theta=\IDLE$,
\begin{align}
    \Phi_n^{\IDLE}(m)
    &=\sum_{k\in\mathcal{P}_n}\psi_{n,k}^R(m)+\max_{\substack{P\subseteq\mathcal{P}_n\\|P|=m}}
    \sum_{k\in P}\big\{\psi_{n,k}^T-\psi_{n,k}^R(m)\big\}.\label{eq:inter-phi-idle}
\end{align}

\subsubsection{Root} \label{sec:dp-root}
Since the root has no parent backhaul, the optimal value of the original problem \eqref{eq:mws} is $\nu_r^\IDLE$.

\subsubsection{Implementation} \label{sec:dp-imp}

Algorithm~\ref{alg:dp} implements the dynamic program through an upward message-passing and a downward schedule decision pass over the IAB tree.
In the upward pass, the IAB nodes are processed from the leaves to the root.
For each feasible number of transmitting panels, Algorithm~\ref{alg:dp-eval} evaluates the optimal sum under modes $T$, $R$, and $\IDLE$.
For a fixed mode and $m$, the panels are sorted in descending order of $\psi_{n,k}^T-\psi_{n,k}^R(m)$ in terms of $k$, and the panels with the largest values are selected.
The number of selected panels is $m$ in modes $T$ and $\IDLE$, and $m-1$ in mode $R$.
For each mode $\theta$ and transmit-panel count $m$, let $\vp_n^\theta(m)$ denote the maximizing panel set $P$ in the corresponding optimization.
After selecting the maximizing count $m_n^\theta$, we store $\vp_n^\theta=\vp_n^\theta(m_n^\theta)$.
This set excludes the reserved parent-backhaul panel in modes $T$ and $R$, so the total transmit-panel count is $m_n^\theta=|\vp_n^\theta|+\mathbf1\{\theta=R\}$.

A leaf IAB computes these values directly from the weights of its access links and parent backhaul link.
An intermediate IAB $n$ first receives $\left(\nu_c^T,\{\nu_c^R(m)\}_{m=0}^{\mnmax},\nu_c^\IDLE\right)$ from every child $c\in\mathcal C_n$.
For the child backhaul link $b_c$, the values associated with its three states are set to $\nu_c^T$ when $b_c^{\mathrm{DL}}$ is activated, $\nu_c^R(m)$ when $b_c^{\mathrm{UL}}$ is activated while $n$ uses $m$ transmitting panels, and $\nu_c^\IDLE$ when $b_c$ is inactive.
These values represent the optimal weighted sums of the descendant subgraph $\mathcal{G}_c$ and are incorporated into $\psi_{n,k}^T$ and $\psi_{n,k}^R(m)$.
Node $n$ then applies Algorithm~\ref{alg:dp-eval} to compute its own optimal values and maximize transmitting-panel assignments.
Each non-root IAB $n$ forwards $\left(\nu_n^T,\{\nu_n^R(m')\}_{m'=0}^{m_{\mathrm{pr}(n)}^{\max}},\nu_n^\IDLE\right)$ to its parent while retaining the maximizing panel assignments required for schedule reconstruction.
Since the root has no parent backhaul link, its mode is fixed to $\IDLE$, and the upward pass terminates with the optimal value of \eqref{eq:mws}.

In the downward pass, the root starts in mode $\IDLE$, and the IAB nodes are processed with each parent preceding its children.
Each IAB retrieves the transmitting-panel assignment $\vp_n^{\theta_n}$ stored for its selected mode and sets $m_n^\star=m_n^{\theta_n}=|\vp_n^{\theta_n}|+\mathbf1\{\theta_n=R\}$, where $m_n^\star$ denotes the optimal total transmit-panel count at node $n$ for the selected mode $\theta_n$.
It then selects links on the corresponding panels as described below.

On a transmitting panel, the effective weight of an access link $\ell$ is $w_\ell$, whereas that of a child downlink $b_c^{\mathrm{DL}}$ is $\nu_c^T-\nu_c^\IDLE$.
The latter is the additional weighted sum obtained by placing $\mathcal{G}_c$ in mode $T$ instead of leaving its parent backhaul idle.
On a receiving panel, the effective weight of an access link $\ell$ is $w_\ell(m_n^\star)$, whereas that of a child uplink $b_c^{\mathrm{UL}}$ is $\nu_c^R(m_n^\star)-\nu_c^\IDLE$.
This difference is the additional weighted sum obtained by placing $\mathcal{G}_c$ in mode $R$ when node $n$ uses $m_n^\star$ transmitting panels.
The link with the largest effective weight is selected on each transmitting panel, while a receiving link is selected only if $m_n^\star\leq\mnmax$ and its largest effective weight is positive; otherwise, the panel remains idle.
The resulting child backhaul decisions determine the modes of the children.
Repeating this procedure until all leaves are processed reconstructs the maximum weighted schedule $\vs^\star$.

\begin{algorithm}[!t]
\small
\caption{Computation of \eqref{eq:dp-nodeval} at IAB $n$}
\label{alg:dp-eval}
\KwInput{$\{\psi_{n,k}^T,\psi_{n,k}^R(m)\}_{k\in\mathcal{P}_n}$, $\{\nu_c^{\IDLE}\}_{c\in\mathcal{C}_n}$, and, if $n\neq r$, $\{R_{b_n^{\mathrm{DL}}}(m,t),w_{b_n^{\mathrm{DL}}}(m)\}_{m=0}^{\mnmax}$ and $\{R_{b_n^{\mathrm{UL}}}(m',t),w_{b_n^{\mathrm{UL}}}(m')\}_{m'=0}^{m_{\mathrm{pr}(n)}^{\max}}$}
\KwOutput{$\nu_n^{\IDLE},\vp_n^{\IDLE}$, and $\nu_n^T,\{\nu_n^R(m')\}_{m'=0}^{m_{\mathrm{pr}(n)}^{\max}},\vp_n^T,\vp_n^R$ if $n\neq r$}

\If{$n\neq r$}{
    $\nu_n^T\gets-\infty$, $\vp_n^T\gets\varnothing$, $\vp_n^R\gets\varnothing$\;
    $\nu_n^R(m')\gets-\infty$ for $m'=0,\ldots,m_{\mathrm{pr}(n)}^{\max}$\;
}

\For{$m\in\{0,\ldots,p_n\}$}{
    \If{$n\neq r$}{
        $\nu_n^{T'}(m)\gets-\infty$, $\nu_n^{R'}(m)\gets-\infty$\;
        $\psi'_{n,k}(m)\eqdef\psi^T_{n,k}-\psi^R_{n,k}(m)$ for $k\in\mathcal{P}_n\setminus\{P_n^1\}$\;
        Sort $p_i\in\mathcal{P}_n\setminus\{P_n^1\}$ by decreasing $\psi'_{n,p_i}(m)$\;
        \If{$m\leq\mnmax$ and $R_{b_n^{\mathrm{DL}}}(m,t)>0$}{
            $\nu_n^{T'}(m)\eqdef w_{b_n^{\mathrm{DL}}}(m)+\sum\limits_{k\in\mathcal{P}_n\setminus\{P_n^1\}}\psi^R_{n,k}(m)+\sum\limits_{i=1}^{m}\psi'_{n,p_i}(m)$\;
            $\vp_n^T(m)\eqdef\{p_1,\ldots,p_m\}$\;
        }
        \If{$m\geq1$}{
            $\nu_n^{R'}(m)\eqdef\sum\limits_{k\in\mathcal{P}_n\setminus\{P_n^1\}}\psi^R_{n,k}(m)+\sum\limits_{i=1}^{m-1}\psi'_{n,p_i}(m)$\;
            $\vp_n^R(m)\eqdef\{p_1,\ldots,p_{m-1}\}$\;
        }
    }
    $\psi''_{n,k}(m)\eqdef\psi^T_{n,k}-\psi^R_{n,k}(m)$ for $k\in\mathcal{P}_n$\;
    Sort $p_i\in\mathcal{P}_n$ by decreasing $\psi''_{n,p_i}(m)$\;
    $\nu_n^{\IDLE'}(m)\eqdef\sum\limits_{k\in\mathcal{P}_n}\psi^R_{n,k}(m)+\sum\limits_{i=1}^{m}\psi''_{n,p_i}(m)$\;
    $\vp_n^{\IDLE}(m)\eqdef\{p_1,\ldots,p_m\}$\;
}

$m_n^{\IDLE}\gets\argmax\limits_{0\leq m\leq p_n}\nu_n^{\IDLE'}(m)$\;
$\nu_n^{\IDLE}\gets\nu_n^{\IDLE'}(m_n^{\IDLE})$, $\vp_n^{\IDLE}\gets\vp_n^{\IDLE}(m_n^{\IDLE})$\;
\If{$n\neq r$}{
    \If{$\max\limits_{0\leq m\leq\mnmax}\nu_n^{T'}(m)>-\infty$}{
        $m_n^T\gets\argmax\limits_{0\leq m\leq\mnmax}\nu_n^{T'}(m)$\;
        $\nu_n^T\gets\nu_n^{T'}(m_n^T)+\sum\limits_{\substack{c\in\mathcal{C}_n\\b_c^{\mathrm{DL}}\in\mathcal{L}_{n,P_n^1}}}\nu_c^{\IDLE}$\;
        $\vp_n^T\gets\vp_n^T(m_n^T)$\;
    }
    \If{$\max\limits_{1\leq m\leq p_n}\nu_n^{R'}(m)>-\infty$}{
        $m_n^R\gets\argmax\limits_{1\leq m\leq p_n}\nu_n^{R'}(m)$\;
        $\vp_n^R\gets\vp_n^R(m_n^R)$\;
        \For{$m'\in\{0,\ldots,m_{\mathrm{pr}(n)}^{\max}\}$ with $R_{b_n^{\mathrm{UL}}}(m',t)>0$}{
            $\nu_n^R(m')\gets w_{b_n^{\mathrm{UL}}}(m')+\nu_n^{R'}(m_n^R)+\sum\limits_{\substack{c\in\mathcal{C}_n\\b_c^{\mathrm{DL}}\in\mathcal{L}_{n,P_n^1}}}\nu_c^{\IDLE}$\;
        }
    }
}
\end{algorithm}

\begin{algorithm}[!t]
\small
\caption{DP for the maximum weighted schedule}
\label{alg:dp}
\KwInput{$\{w_\ell(m),R_\ell(m,t)\}_{\ell\in\mathcal{L}}$}
\KwOutput{$\vs^\star$}

\textit{Phase 1: Upward weight computation from leaves to the root}\;

\For{leaf IAB $n$ of $\mathcal{G}$}{
    Evaluate $\psi_{n,k}^T$ and $\psi_{n,k}^R(m)$ by \eqref{eq:leaf-psiT}, \eqref{eq:leaf-psiR} over available access links, for $0\leq m\leq\mnmax$\;
    Set $\psi_{n,k}^R(m)\gets0$ for $\mnmax<m\leq p_n$ and $k\in\mathcal{P}_n$\;
    Run Algorithm~\ref{alg:dp-eval} and retain its outputs at $n$\;
    \If{$n\neq r$}{
        Send $\left(\nu_n^T,\{\nu_n^R(m')\}_{m'=0}^{m_{\mathrm{pr}(n)}^{\max}},\nu_n^{\IDLE}\right)$ to $\mathrm{pr}(n)$\;
    }
}

\For{non-leaf IAB $n$, children before parents}{
    Receive $\big(\nu_c^T,\{\nu_c^R(m)\}_{m=0}^{\mnmax},\nu_c^{\IDLE}\big)$ from every $c\in\mathcal{C}_n$\;
    \For{$c\in\mathcal{C}_n$}{
        $\tau_{b_c^{\mathrm{DL}}}\eqdef\nu_c^T$, $\rho_{b_c^{\mathrm{UL}}}(m)\eqdef\nu_c^R(m)$ for $0\leq m\leq\mnmax$\;
        $\phi_{b_c^{\mathrm{DL}}}=\phi_{b_c^{\mathrm{UL}}}\eqdef\nu_c^{\IDLE}$\;
    }
    Get $\psi_{n,k}^T$ and $\psi_{n,k}^R(m)$ by \eqref{eq:inter-T}, \eqref{eq:inter-R}, for $0\leq m\leq\mnmax$\;
    Set $\psi_{n,k}^R(m)\gets B_{n,k}$ for $\mnmax<m\leq p_n$ and $k\in\mathcal{P}_n$\;
    Run Algorithm~\ref{alg:dp-eval} and retain its outputs at $n$\;
    \If{$n\neq r$}{
        Send $\left(\nu_n^T,\{\nu_n^R(m')\}_{m'=0}^{m_{\mathrm{pr}(n)}^{\max}},\nu_n^{\IDLE}\right)$ to $\mathrm{pr}(n)$\;
    }
}

\textit{Phase 2: Downward schedule selection from the root to the leaves}\;

$\theta_r\gets\IDLE$, $\vs^\star=[s_\ell^\star]_{\ell\in\mathcal{L}}\gets\mathbf{0}$\;

\For{$n\in\mathcal{B}$, parents before children}{
    \If{$n\neq r$}{
        \lIf{$s_{b_n^{\mathrm{DL}}}^\star=1$}{$\theta_n\gets T$}
        \lElseIf{$s_{b_n^{\mathrm{UL}}}^\star=1$}{$\theta_n\gets R$}
        \lElse{$\theta_n\gets\IDLE$}
    }
    $m_n^\star\gets|\vp_n^{\theta_n}|+\mathbf{1}\{\theta_n=R\}$, $\mathcal{P}_n'\gets\mathcal{P}_n$\;

    \If{$n\neq r$ and $\theta_n\in\{T,R\}$}{
        $\mathcal{P}_n'\gets\mathcal{P}_n\setminus\{P_n^1\}$\;
    }
    $w_\ell^{\mathrm{TX}}\gets-\infty$, $w_\ell^{\mathrm{RX}}\gets-\infty$ for $\ell\in\mathcal{L}_n\cap\mathcal{L}(\mathcal{G}_n)$\;

    \For{$k\in\vp_n^{\theta_n}\cap\mathcal{P}_n'$}{
        \For{$\ell\in\mathcal{L}_{n,k}^{\mathrm{TX}}\cap\mathcal{L}(\mathcal{G}_n)$ with $R_\ell(0,t)>0$}{
            \lIf{$\ell\in\mathcal{L}_{\mathrm{AC}}$}{$w_\ell^{\mathrm{TX}}\gets w_\ell$}
            \uElseIf{$\ell=b_c^{\mathrm{DL}}$, $c\in\mathcal{C}_n$}{
                $w_\ell^{\mathrm{TX}}\gets\nu_c^T-\nu_c^{\IDLE}$\;
            }
        }
        $\ell_{n,k}^{\mathrm{TX},\star}\gets\argmax_{\ell\in\mathcal{L}_{n,k}^{\mathrm{TX}}\cap\mathcal{L}(\mathcal{G}_n)}w_\ell^{\mathrm{TX}}$\;
        $s_{\ell_{n,k}^{\mathrm{TX},\star}}^\star\gets1$\;
    }

    \If{$m_n^\star\leq\mnmax$}{
        \For{$k\in\mathcal{P}_n'\setminus\vp_n^{\theta_n}$}{
            \For{$\ell\in\mathcal{L}_{n,k}^{\mathrm{RX}}\cap\mathcal{L}(\mathcal{G}_n)$ with $R_\ell(m_n^\star,t)>0$}{
                \lIf{$\ell\in\mathcal{L}_{\mathrm{AC}}$}{$w_\ell^{\mathrm{RX}}\gets w_\ell(m_n^\star)$}
                \uElseIf{$\ell=b_c^{\mathrm{UL}}$, $c\in\mathcal{C}_n$}{
                    $w_\ell^{\mathrm{RX}}\gets\nu_c^R(m_n^\star)-\nu_c^{\IDLE}$\;
                }
            }
            \If{$\mathcal{L}_{n,k}^{\mathrm{RX}}\cap\mathcal{L}(\mathcal{G}_n)\neq\varnothing$}{
                $\ell_{n,k}^{\mathrm{RX},\star}\gets\argmax_{\ell\in\mathcal{L}_{n,k}^{\mathrm{RX}}\cap\mathcal{L}(\mathcal{G}_n)}w_\ell^{\mathrm{RX}}$\;
                \lIf{$w_{\ell_{n,k}^{\mathrm{RX},\star}}^{\mathrm{RX}}>0$}{$s_{\ell_{n,k}^{\mathrm{RX},\star}}^\star\gets1$}
            }
        }
    }
}
\end{algorithm}

\subsubsection{Computational Complexity} \label{sec:dp-complexity}

At node $n$, the values received from each child $c\in\mathcal{C}_n$ are used to form the effective link values $\tau_\ell$, $\rho_\ell(m)$, and $\phi_\ell$ for the child backhaul links.
Since these values are evaluated over at most $p_n+1$ possible transmitting-panel counts, this step costs $\mathcal{O}(p_n|\mathcal{C}_n|)$.

Next, for each count $m$, each panel $k\in\mathcal{P}_n$ selects its best transmitting or receiving link by taking a maximum over the links in $\mathcal{L}_{n,k}$.
Over all panels and all counts, this requires $\mathcal{O}(p_n|\mathcal{L}_n|)$ operations.
For each count, Algorithm~\ref{alg:dp-eval} then sorts the $p_n$ panel gains to choose the best set of transmitting panels, which costs $\mathcal{O}(p_n\log(p_n))$ per count and hence $\mathcal{O}(p_n^2\log(p_n))$ in total.

Therefore, the computation at node $n$ costs $\mathcal{O}(p_n|\mathcal{L}_n|+p_n^2\log(p_n)+p_{\mathrm{pr}(n)})$, where the $\mathcal{O}(p_n|\mathcal{C}_n|)$ term is absorbed because each child backhaul link is included in $\mathcal{L}_n$.
If $p_n\leq p$ for all nodes and $\sum_n|\mathcal{L}_n|\leq 2|\mathcal{L}|$, the upward pass costs $\mathcal{O}(p|\mathcal{L}|+Bp^2\log(p))$.
The downward pass selects links on the stored panel assignments and propagates the corresponding modes to the children, so it costs $\mathcal{O}(|\mathcal{L}|+Bp)$ and is dominated by the upward pass.
Thus, the total complexity is $\mathcal{O}(p|\mathcal{L}|+Bp^2\log(p))$, which is almost linear in the number of links.

\subsection{Stability region of the IAB network}
\begin{definition} \label{def:stability}
    The system is said to be stable under a scheduling policy if and only if
    \begin{align}
        \limsup_{t\to\infty}\frac{1}{t}\sum_{\tau=0}^{t-1}\mathbb{E}\left[\boldsymbol Q(\tau)\right]<\infty.
    \end{align}
\end{definition}

\begin{definition} \label{def:region}
    The system is called stabilizable if and only if there exists a stationary scheduling policy under which the system is stable.
\end{definition}

For each $\{\mathrm{SNR}_\ell\}_{\ell\in\mathcal{L}}$, define the set of feasible service rate vectors as
\begin{align}
    \mathcal{C}\bigl(\{\mathrm{SNR}_\ell\}\bigr) \eqdef \left\{\left[R_\ell\bigl(m_\ell(\vs)\bigr)s_\ell\right]_{\ell\in\mathcal{L}} : \vs\in \mathcal{S}\bigl(\{\mathrm{SNR}_\ell\}\bigr)\right\}. \label{eq:instantaneous-service-region}
\end{align}
Here, $\operatorname{Conv}$ denotes the convex hull.
Accordingly, the average service region is
\begin{align}
    \overline{\mathcal{C}} \eqdef \left\{\mathbb{E}\left[\vc\bigl(\{\mathrm{SNR}_\ell\}\bigr)\right] : \vc\bigl(\{\mathrm{SNR}_\ell\}\bigr) \in \operatorname{Conv}\left(\mathcal{C}\bigl(\{\mathrm{SNR}_\ell\}\bigr)\right)\right\},    \label{eq:average-service-region}
\end{align}
where the expectation is taken over the distribution of $\{\mathrm{SNR}_\ell\}_{\ell\in\mathcal{L}}$.

For an arrival rate vector $\vlambda=[\lambda_f]_{f\in\mathcal{F}}$, let $\lambda_\ell\eqdef\sum_{f\in\mathcal{F}_\ell}\lambda_f$ denote the aggregate load induced on link $\ell$.
The stability region is
\begin{align}
    \Lambda \eqdef \left\{\vlambda\in\mathbb{R}_+^{|\mathcal{F}|} : \exists\, \overline{\vc}\in\overline{\mathcal{C}} \text{ such that } \lambda_\ell \leq \overline{c}_\ell, \ \forall\ell\in\mathcal{L}\right\}. \label{eq:stability-region}
\end{align}

\begin{theorem} \label{thm:region}
If $\vlambda \notin \Lambda$, the system cannot be stable under any scheduling policy.
For a $\vlambda$, if there exists a $\delta > 0$ such that $\vlambda + \delta \mathbf{1} \in \Lambda$, where $\mathbf{1} = [1]_{f \in \mathcal{F}}$, there exists at least one stationary scheduling policy that can stabilize the system.
We call such a $\vlambda$ an interior point of $\Lambda$ and denote the interior of $\Lambda$ as $\operatorname{int}(\Lambda)$.
\end{theorem}

\begin{proof}
The proof follows from the standard proof of Theorem 5 of \cite{neely2003dynamic}.
\end{proof}

\subsection{Global Back-Pressure Policy} \label{sec:global-bp}

The global back-pressure policy instantiates the maximum weighted schedule \eqref{eq:mws} with the differential-backlog weight.
On a downlink, $\ell \in \mathcal{L}_{\mathrm{AC}}\cap\mathcal{L}_{\mathrm{DL}}$,  since the UE is the end point of the flow and keeps no queue, the weight of an access link incident to node $n$ is
\begin{align} \label{eq:bp-access}
    w_\ell^{\mathrm{bp}}(t) = R_\ell(t) \max_{f \in \mathcal{F}_\ell} q_n^f(t).
\end{align}
For an uplink access link received at IAB $n$, $\ell\in\mathcal{L}_{\mathrm{AC}}\cap\mathcal{L}_{\mathrm{UL}}$,
\begin{align} \label{eq:bp-access-ul}
    w_\ell^{\mathrm{bp}}(m,t) = R_\ell(m,t) \max_{f \in \mathcal{F}_\ell} \big(q_{A(\ell)}^f(t)-q_n^f(t)\big)^+.
\end{align}
For a backhaul link between two IABs, differential weights are defined following the direction of flow as
\begin{align} \label{eq:bp-bh-dl}
    w_{b_n^{\mathrm{DL}}}^{\mathrm{bp}}(m,t) &= R_{b_n^{\mathrm{DL}}}(m,t) \max_{f \in \mathcal{F}_{b_n^{\mathrm{DL}}}} \big(q_{\mathrm{pr}(n)}^f(t) - q_n^f(t)\big)^+, \\
    \label{eq:bp-bh-ul}
    w_{b_n^{\mathrm{UL}}}^{\mathrm{bp}}(m,t) &= R_{b_n^{\mathrm{UL}}}(m,t) \max_{f \in \mathcal{F}_{b_n^{\mathrm{UL}}}} \big(q_n^f(t) - q_{\mathrm{pr}(n)}^f(t)\big)^+.
\end{align}
Hence, under the global back-pressure policy, the schedule is determined by optimizing the following objective through Algorithms 1 and 2.
\begin{align} \label{eq:bp-schedule}
    \vs(t) \eqdef \argmax_{\vs \in \mathcal{S}} \sum_{\ell \in \mathcal{L}} w_\ell^{\mathrm{bp}}\bigl(m_\ell(\vs,t),t\bigr)\, s_\ell.
\end{align}
The scheduling policy sends only the packet corresponding to flow $f^\star \in \argmax_{f \in \mathcal{F}_\ell}\big(q_{I(A(\ell))}^f(t)-q_{I(D(\ell))}^f(t)\big)$ for a backhaul link $\ell$.

To discuss the stability of the back-pressure policy, the following assumptions are made.
\begin{assumption}
\leavevmode
\begin{enumerate}
    \item The arrival process $\{\lambda_f(t)\}_{t=0}^{\infty}$ of each flow $f\in\mathcal{F}$ is a stationary and ergodic random process and has a finite second moment, $\mathbb{E}\left[\left(\lambda_f(t)\right)^2\right] < \infty$.
    \item The SNR random processes $\{\mathrm{SNR}_\ell(t)\}_{t=0}^{\infty}$ where $\ell \in \mathcal{L}$ are independent across times.
\end{enumerate}
\end{assumption}

\begin{theorem} \label{thm:bp}
The global back-pressure policy is stable for every arrival-rate vector in the interior of $\Lambda$.
\end{theorem}

\begin{proof}
Refer to \cite{tassiulas1990stability, georgiadis2006resource} for the proof.
\end{proof}

\section{Reduced-overhead Policy}
\label{sec:reduced-overhead-mw}

\begin{figure*}[!t]
    \centering
    \includegraphics[
    width=\textwidth,
    height=0.42\textheight,
    keepaspectratio
]{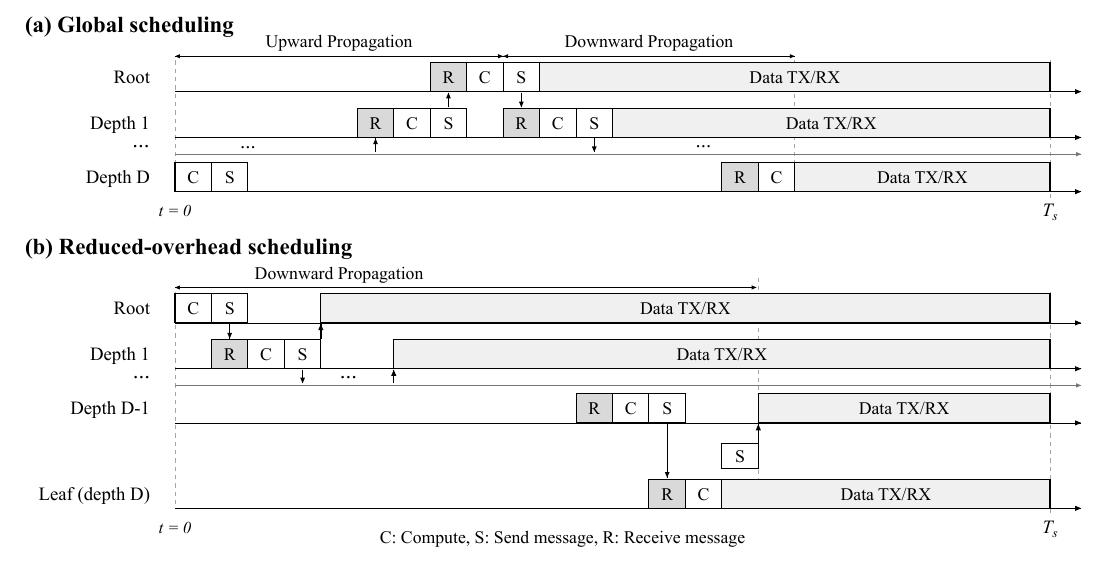}
    \caption{Timing diagram of global and reduced-overhead policies.}
    \label{fig:timing-diagram}
\end{figure*}

The global policy requires subtree scheduling values to be collected through an upward pass before scheduling decisions can propagate from the root.
To avoid waiting for this information to be collected across the network, we consider reduced overhead policies that select transmissions using each IAB's queues, incident channel states, and neighbour information.
Each non-root IAB selects its schedule upon receiving its parent's backhaul decision, without waiting for its children's decisions.
Once a receiving child has selected its schedule, it reports its transmit-panel count to its parent for backhaul rate and MCS selection.
As soon as this coordination between the parent and child is complete, packet transmission begins on the selected link, even while scheduling continues elsewhere in the network.
The timing diagram of the reduced overhead policy is described in Fig. \ref{fig:timing-diagram}.
We develop FROM to implement this reduced overhead scheduling approach.

\subsection{Floor-Based Reduced Overhead MaxWeight}
\label{sec:FROM}

A parent selects an outgoing downlink backhaul before the receiving child determines its simultaneous transmissions, so the parent cannot yet evaluate the backhaul rate exactly.
FROM instead uses a fixed minimum backhaul rate.
This prevents overestimating the backhaul weight and giving it excessive priority over competing links.
The backhaul reception limit $\mnmax$ in \eqref{eq:node-tx-limit} is computed from the SNRs without outage and remains fixed across slots.
Since the backhaul rate is non-increasing in $m$, its minimum rate under this limit is
\begin{equation}
\label{eq:FROM-floor-rate}
R_{b_n^{\mathrm{DL}}}^{-}
\eqdef R_{b_n^{\mathrm{DL}}}^{\mathrm{BH}}
\bigl(\mnmax\bigr).
\end{equation}

FROM weights each link by its aggregate queue length,
\begin{equation}
\label{eq:FROM-queue}
Q_\ell(t)\eqdef
\begin{cases}
\displaystyle\sum_{f\in\mathcal F_\ell}q_{A(\ell)}^f(t),
&\ell\in\mathcal L_{\mathrm{AC}}\cap\mathcal L_{\mathrm{UL}},\\[2pt]
\displaystyle\sum_{f\in\mathcal F_\ell}q_{I(A(\ell))}^f(t),
&\text{otherwise}.
\end{cases}
\end{equation}

\subsubsection{Reduced Overhead Scheduling}
\label{sec:FROM-reduced overhead-scheduling}

At the beginning of slot $t$, each non-root IAB $c$ sends $Q_{b_c^{\mathrm{UL}}}(t)$ to its parent for evaluating the uplink backhaul.
Each IAB then selects its reduced overhead schedule under the inherited mode in \eqref{eq:bn-mode}, starting with $\theta_r(t)=\IDLE$.
Let $\mathcal S_n^\theta(t)$ contain the reduced overhead schedules satisfying the physical constraints at $n$, the inherited mode $\theta$, and
\begin{equation}
\label{eq:FROM-feasible-set}
\begin{aligned}
s_\ell(t)&\le\mathbf1\{Q_\ell(t)>0\},
&&\ell\in\mathcal L_n,\\
\left\lceil R_{b_c^{\mathrm{DL}}}^{-}\right\rceil
s_{b_c^{\mathrm{DL}}}(t)&\le Q_{b_c^{\mathrm{DL}}}(t),
&&c\in\mathcal C_n.
\end{aligned}
\end{equation}
The second inequality is the backhaul queue-eligibility constraint: a downlink backhaul can be activated only after enough traffic has accumulated to use its minimum rate.
This limits frequent backhaul activation with little queued traffic and leaves the child's parent-facing panel available for other links.

Each IAB $n$ selects
\begin{align}
\vs_n^{\mathrm{FROM}}(t)
&\eqdef\argmax_{\vs_n\in\mathcal S_n^{\theta_n(t)}(t)}
\Bigg\{\sum_{c\in\mathcal C_n}
Q_{b_c^{\mathrm{DL}}}R_{b_c^{\mathrm{DL}}}^{-}
s_{b_c^{\mathrm{DL}}} +
\nonumber\\
&\quad\sum_{\ell\in\mathcal L_n^{\mathrm{RX}}
\cup(\mathcal L_n\cap\mathcal L_{\mathrm{AC}})}
Q_\ell R_\ell\!\bigl(m_\ell(\vs_n), t\bigr)s_\ell
\Bigg\}.
\label{eq:FROM-reduced-overhead-mw}
\end{align}
Backhaul receptions at $n$ use their exact rates because node $n$ can determine its transmit-panel count and the resulting self-interference, whereas downlink backhauls to children use the minimum rates in \eqref{eq:FROM-floor-rate}.

For each feasible transmit-panel count $m$, we compute the maximum TX weight and the maximum RX-or-idle weight on each available panel, as in \eqref{eq:leaf-psiT} and \eqref{eq:leaf-psiR}.
The candidate links include both access links and child backhauls satisfying \eqref{eq:FROM-feasible-set}, with their weights given by \eqref{eq:FROM-reduced-overhead-mw}.
The transmitting panels are selected as in \eqref{eq:leaf-T}--\eqref{eq:leaf-phi}, and the resulting candidates are compared using the full objective in \eqref{eq:FROM-reduced-overhead-mw}.
After IAB $n$ selects its schedule, $n$ sends its decision and queue information to its children.
Then, the children make their own scheduling decisions and send their transmit-panel count to $n$.
Based on the transmit-panel count of the children, $n$ chooses the corresponding MCS and transmits packets to its children.
Algorithm~\ref{alg:FROM} summarises the procedure.

\begin{algorithm}[!t]
\footnotesize
\caption{FROM scheduling policy}
\label{alg:FROM}
\KwInput{$\{Q_\ell(t),\snr_\ell(t)\}_{\ell\in\mathcal L_n}$,
$\{R_{b_c^{\mathrm{DL}}}^{-}\}_{c\in\mathcal C_n}$}
\KwOutput{$\vs^{\mathrm{FROM}}(t)$}
\For{non-root IAB $c$ in parallel}{
    Send $Q_{b_c^{\mathrm{UL}}}(t)$ to $\mathrm{pr}(c)$\;
}
$\theta_r(t)\gets\IDLE$\;
\For{$n\in\mathcal B$, parents before children}{
    $\mathcal P_n'\gets\mathcal P_n$\;
    \lIf{$\theta_n(t)\neq\IDLE$}{$\mathcal P_n'\gets\mathcal P_n\setminus\{P_n^1\}$}
    \For{feasible transmit-panel count $m$}{
        Compute $\psi_{n,k}^T,\psi_{n,k}^R(m)$ over links eligible under \eqref{eq:FROM-feasible-set}, using the weights in \eqref{eq:FROM-reduced-overhead-mw}\;
        Select $m-\mathbf1\{\theta_n(t)=R\}$ panels in $\mathcal P_n'$ with the largest $\psi_{n,k}^T-\psi_{n,k}^R(m)$\;
        Form $\vs_n^{(m)}(t)$ from the maximising TX and RX/idle choices, retaining the inherited mode\;
    }
    Select $\vs_n^{\mathrm{FROM}}(t)$ maximising \eqref{eq:FROM-reduced-overhead-mw} among the candidates\;
    \For{$c\in\mathcal C_n$}{
        Set $\theta_c(t)$ by \eqref{eq:bn-mode} and send it to $c$\;
        \lIf{$\theta_c(t)=T$}{also send $Q_{b_c^{\mathrm{DL}}}(t)$ to $c$}
    }
}
\For{$n\neq r$ with $\theta_n(t)=T$ in parallel}{
    Send $m_n(\vs_n^{\mathrm{FROM}}(t),t)$ to $\mathrm{pr}(n)$\;
}
Execute $\vs^{\mathrm{FROM}}(t)$ using the physical link rates\;
\end{algorithm}

\subsection{Stability Region of Reduced Overhead Policies}
\label{sec:reduced-overhead-policy-region}

We consider only downlink traffic in the stability analysis in this and the following subsections.
Downlink traffic alone is sufficient to analyse the scheduling dependence addressed by FROM, since a parent must select a backhaul before knowing the receiving child's simultaneous transmissions and the resulting backhaul rate.

\begin{assumption}
\label{assump:reduced-overhead-independence}
For each $f \in \mathcal F$, the exogenous arrival process $\{\lambda_f(t)\}_{t=0}^\infty$ is an i.i.d. sequence of random variables with $\mathbb{E}\left[(\lambda_f(t))^2\right] < \infty$.
Let $\snr_n(t) \eqdef [\snr_\ell (t)]_{\ell \in \mathcal{L}_n}$.
For each $n \in \mathcal{B}$, $\{\snr_n(t)\}_{t=0}^\infty$ is an i.i.d. sequence of random variables.
The SNR processes are independent across links and times and are independent of the arrival processes.
\end{assumption}

A deterministic reduced overhead rule $d_n(\theta_n,\snr_n)$ selects $n$'s outgoing downlinks using only the backhaul decision $\theta_n$ received from its parent and its channel states $\snr_n$.
Let $d=(d_n)_{n\in\mathcal B}$ and let $\vs^d(\snr)$ be the schedule obtained by applying these rules from the root to the leaves, with $\theta_r=\IDLE$.
We define the class of feasible reduced overhead rules by
\begin{equation}
\label{eq:FROM-reduced-overhead-rule-class}
\mathfrak D_{\mathrm{ro}}
\eqdef
\left\{
d:\vs^d(\snr)\in\mathcal S(\snr),
\ \forall\snr
\right\}.
\end{equation}
Write $\vs^d(t)=\vs^d(\snr(t))$ and let $\snr_n^{\mathrm{out}}(t)=(\snr_\ell(t))_{\ell\in\mathcal L_n^{\mathrm{TX}}\cap\mathcal L_{\mathrm{DL}}}$ denote the channel states of $n$'s outgoing downlinks.
The parent of $n$ chooses $s_{b_n^{\mathrm{DL}}}^d(t)$ without using $\snr_n^{\mathrm{out}}(t)$.
Under Assumption~\ref{assump:reduced-overhead-independence}, every $d\in\mathfrak D_{\mathrm{ro}}$ therefore satisfies
\begin{equation}
\label{eq:FROM-reduced-overhead-independence}
\mathbb E\!\left[s_{b_n^{\mathrm{DL}}}^d(t)
\mid\snr_n^{\mathrm{out}}(t)\right]
=\mathbb E\!\left[s_{b_n^{\mathrm{DL}}}^d(t)\right],
\quad n\ne r.
\end{equation}
The corresponding offered service vector is
\begin{equation}
\label{eq:FROM-rule-service}
\vc^d(t)=
\left(
R_\ell\bigl(m_\ell(\vs^d(t),t),t\bigr)s_\ell^d(t)
\right)_{\ell\in\mathcal L_{\mathrm{DL}}}.
\end{equation}
The mean service region consists of all convex combinations of the expected service vectors generated by feasible reduced overhead rules:
\begin{equation}
\label{eq:FROM-reduced-overhead-service-region}
\overline{\mathcal C}_{\mathrm{ro}}
\eqdef\operatorname{Conv}
\left\{\mathbb E[\vc^d(t)]:d\in\mathfrak D_{\mathrm{ro}}\right\}.
\end{equation}
The corresponding arrival region consists of rates whose load on each downlink does not exceed its mean service:
\begin{equation}
\label{eq:FROM-reduced-overhead-arrival-region}
\Lambda_{\mathrm{ro}}\eqdef\left\{\vlambda\in\mathbb R_+^{|\mathcal F_{\mathrm{DL}}|}\;\middle|\;\exists\,\overline{\vc}\in\overline{\mathcal C}_{\mathrm{ro}},\ \lambda_\ell\le\overline c_\ell,\ \forall\ell\in\mathcal L_{\mathrm{DL}}\right\}.
\end{equation}

\begin{theorem}
\label{thm:reduced-overhead-policy-region}
\begin{enumerate}
\item If $\vlambda\notin\Lambda_{\mathrm{ro}}$, no stationary reduced overhead policy stabilizes the network.
\item If $\vlambda+\varepsilon\boldsymbol1\in\Lambda_{\mathrm{ro}}$ for some $\varepsilon>0$, a randomized stationary reduced overhead policy stabilizes the network, where $\boldsymbol1$ is indexed by downlink flows.
\end{enumerate}
\end{theorem}
\begin{proof}
See Appendix~\ref{app:reduced-overhead-policy-region} of the supplementary material.
\end{proof}

\begin{proposition}
\label{prop:reduced overhead-global-region}
The relationship between the global stability region and the reduced overhead stability region is
\begin{equation}
\label{eq:reduced overhead-global-region}
\frac12\Lambda\subseteq\Lambda_{\mathrm{ro}}\subseteq\Lambda.
\end{equation}
\end{proposition}
\begin{proof}
See Appendix~\ref{app:local-global-region} of the supplementary material.
\end{proof}

\subsection{Stability Region of FROM}
\label{sec:reduced overhead-stability}

Set
\begin{equation}
\label{eq:FROM-ceiling-rate}
R_{b_n^{\mathrm{DL}}}^{+}
\eqdef R_{b_n^{\mathrm{DL}}}^{\mathrm{BH}}(0).
\end{equation}
Define
\begin{equation}
\label{eq:FROM-beta}
\beta\eqdef
\min_{n\in\mathcal B\setminus\{r\}}
\frac{R_{b_n^{\mathrm{DL}}}^{-}}
     {R_{b_n^{\mathrm{DL}}}^{+}},
\end{equation}
with $\beta=1$ if there is no backhaul.

\begin{theorem}
\label{thm:FROM-stability}
FROM stabilizes $\vlambda\in\operatorname{int}(\beta\Lambda_{\mathrm{ro}})$.
\end{theorem}
\begin{proof}
See Appendix~\ref{app:FROM-stability} of the supplementary material.
\end{proof}

\begin{corollary}
\label{cor:FROM-constant-channel}
Suppose that all link SNRs are constant and no outages occur.
If $\mathrm{INR}=0$ or $\mathrm{INR}=\infty$, then $\beta=1$ and
\begin{equation}
\label{eq:FROM-endpoint-region}
\operatorname{int}(\Lambda)
=\operatorname{int}(\Lambda_{\mathrm{ro}}).
\end{equation}
\end{corollary}

\section{Numerical Analysis} \label{sec:numerical-results}

\begin{table}[t]
\centering
\caption{Simulation Setup}
\label{tab:simulation-parameters}
\footnotesize
\setlength{\tabcolsep}{4pt}
\renewcommand{\arraystretch}{1.05}
\begin{tabularx}{\columnwidth}{
    @{}>{\raggedright\arraybackslash}X
    >{\centering\arraybackslash}p{0.36\columnwidth}@{}
}
\hline
\textbf{Parameters} & \textbf{Values} \\
\hline
Carrier frequency, $f_c$ (GHz)
& $23$ \\
Bandwidth, $\mathrm{BW}$ (GHz)
& $1$ \\
Slot duration, $T_s$ ($\mu$s)
& $125$ \\
Packet size, $L_{\mathrm{pkt}}$ (kb)
& $100$ \\
Noise power spectral density (dBm/Hz)
& $-174$ \\
BH receiver noise figure (dB)
& $5$ \\
AC receiver noise figure (dB)
& $7$ \\
BH transmit power (dBm)
& $24$ \\
BH beamforming gain (dB)
& $36$ \\
AC-DL transmit power (dBm)
& $24$ \\
AC-DL beamforming gain (dB)
& $30$ \\
AC-UL transmit power (dBm)
& $23$ \\
AC-UL beamforming gain (dB)
& $23$ \\
Backhaul distance (m)
& $[400,600]$ \\
Access distance (m)
& $[10,200]$ \\
Total number of IABs, including the donor
& $11$ \\
Number of panels per IAB
& $4$ \\
Number of UEs per IAB
& $[0,16]$ \\
Number of UEs per panel
& $[0,4]$ \\
Rician $K$-factor (dB)
& $13$ \\
Access outage probability
& $0.1$ \\
Backhaul outage probability
& $0.01$ \\
Minimum spectral efficiency (bit/s/Hz)
& $0.2344$ \\
DL/UL traffic ratio
& $90\%/10\%$ \\
Simulation length (slots)
& $30000$ \\
\hline
\end{tabularx}

\smallskip
\begin{minipage}{\columnwidth}
\footnotesize
\end{minipage}
\end{table}

To assess the proposed full-duplex IAB scheduling policy, we conducted numerical analysis under the full-duplex system illustrated in Fig.~\ref{fig:iab-system}\subref{fig:iab-topology}.
The network consists of one donor and ten relay IABs, with a depth of three.
Each IAB is equipped with four panels.
The number of UEs associated with each panel is drawn uniformly from $\{0, \ldots,4\}$.
The distances of IAB-IAB and IAB-UE are drawn uniformly from $[400,600]$~m and $[10,200]$~m, respectively.

Following \cite{gopalam2022distributed}, we set the carrier frequency as 23~GHz, and the bandwidth as 1~GHz.
IABs and UEs transmit at 24~dBm and 23~dBm, respectively.
The beamforming gains of backhaul links are 36~dB, 30~dB for access downlinks, and 23~dB for access uplinks.
Receiver noise figures are 5~dB for backhaul links and 7~dB for access links in both directions.
At a link distance of $d$ meters, the path loss is $32.4+21\log_{10}d+20\log_{10}23$~dB.
The minimum supported spectral efficiency is set to $0.2344$~bit/s/Hz, corresponding to MCS index 0 in \cite{3gpp38214}, and the corresponding minimum SINR, $\gamma_{\min}=2^{0.2344}-1 \approx -7.53$~dB.

Access links experience Rician fading with a $K$-factor of 13~dB.
Following \cite{gapeyenko2017temporal}, due to tracking errors or obstacles, outages occur independently in each slot, with probabilities $0.1$ for access links and $0.01$ for backhaul links.

Each UE has one downlink flow from the donor and one uplink flow to the donor, both following the fixed tree path.
Packet arrivals follow independent Poisson processes across flows and slots.
The load of each UE, $\lambda$, denotes the combined downlink and uplink arrival rate, and the downlink and uplink flows receive $0.9\lambda$ and $0.1\lambda$, respectively.
Each packet contains 100~kb, and each slot lasts $125~\mu$s.

\begin{figure}[t]
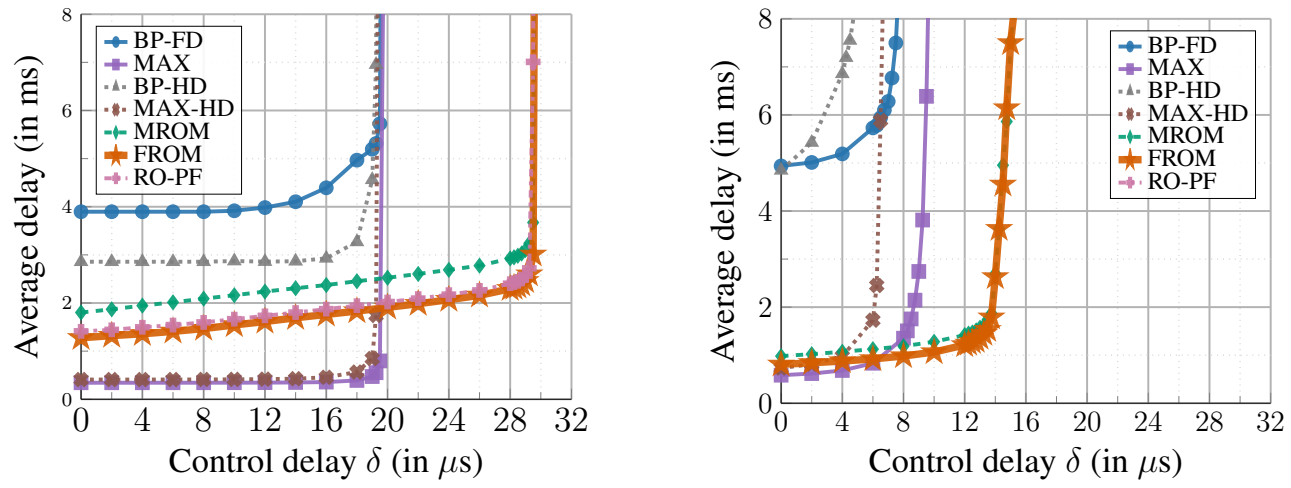

\centering
\subfloat[$\lambda=40$ Mb/s]{%
    \resizebox{0.48\columnwidth}{!}{%
        \input{fig/control/control_lam40}%
    }%
    \label{fig:control_lam40}%
}\hfill
\subfloat[$\lambda=120$ Mb/s]{%
    \resizebox{0.48\columnwidth}{!}{%
        \input{fig/control/control_lam120}%
    }%
    \label{fig:control_lam120}%
}
\caption{Average delay versus control delay $\delta$ at $\inr =0$ dB.}
\label{fig:control_delay}
\end{figure}

To represent the control delay required for exchanging messages across the tree, we will denote the node-to-node control delay as $\delta$.
Control signaling uses resources separate from data transmission, and packet transmission is halted during time dedicated to control-message exchange.
The computation time for scheduling is omitted for this numerical analysis to concentrate on the communication delay for control.
Let $h$ denote the maximum depth of the tree and $h_n$ the depth of node $n$, with the root at depth zero.
For a global scheduling policy, the upward message propagation takes $h\delta$.
The downward propagation takes $h_n\delta$ since it needs to reach the depth $h_n$.
Finally, if a node $n$ has children, it takes $\delta$ to hand over control messages to the children.
Thus, the control delay for the global scheduling policy is $t_n^c = (h+h_n+\chi_n)\delta$, where $\chi_n=1$ if $n$ has IAB children and $\chi_n=0$ otherwise.
For a reduced overhead scheduling policy, it takes $h_n \delta$ for control messages to reach the node $n$.
To synchronize communication settings like MCS, node $n$ and its children should pass its scheduling to each other.
Hence, for a reduced overhead scheduling policy, the control delay is $t_n^c = (h_n + 2\chi_n)\delta$.
The available time for packet transfer on the links of $n$ is then scaled by $1-t_n^c/T_{\mathrm{s}}$.
We omit the delay incurred when non-root IABs simultaneously send their uplink queue information to their parents at the beginning of each slot, as this exchange is common to both global and reduced-overhead scheduling.
We set $\delta = 0$ except in the experiment shown in Fig. \ref{fig:control_delay}.

We consider the following seven scheduling policies for comparison:
\begin{itemize}
    \item \textbf{BP-FD:} This policy uses global information to maximize $\sum_{\ell\in\mathcal L}w_\ell^{\mathrm{bp}}(m,t)s_\ell(t)$ under the FD scheduling constraints, where $w_\ell^{\mathrm{bp}}(t)$ is the back-pressure weight defined in Section \ref{sec:global-bp}.

    \item \textbf{MAX:} This policy uses global information to maximize $\sum_{\ell\in\mathcal L}Q_\ell(t)R_\ell(m,t)s_\ell(t)$ under the FD scheduling constraints proposed in Section \ref{sec:scheduling}.

    \item \textbf{MAX-HD} \cite{gopalam2022distributed}: This policy uses the same scheduling objective rule as MAX, but prohibits simultaneous transmission and reception at each IAB and only allows a half-duplex operation.

    \item \textbf{BP-HD} \cite{gopalam2022distributed}: This policy uses the same scheduling objective rule as BP-FD, but prohibits simultaneous transmission and reception at each IAB and only allows a half-duplex operation.
    
    \item \textbf{FROM:} The proposed reduced overhead policy evaluates children's downlink backhauls using their minimum admissible rates proposed in Section \ref{sec:reduced-overhead-mw}.

    \item \textbf{RO-PF}: This reduced overhead proportional fair policy uses the same rate estimates as FROM but maximizes the sum of scheduled link rates normalized by their average achieved throughputs.
    This normalization favors links with lower past throughput, promoting fairness among links.

    \item \textbf{MROM:} This Maximum-based Reduced-Overhead MaxWeight replaces the minimum backhaul rates in FROM's scheduling objective and queue constraint with $R^+_{b_n^{\mathrm{DL}}}$.

\end{itemize}

\begin{figure}[t]
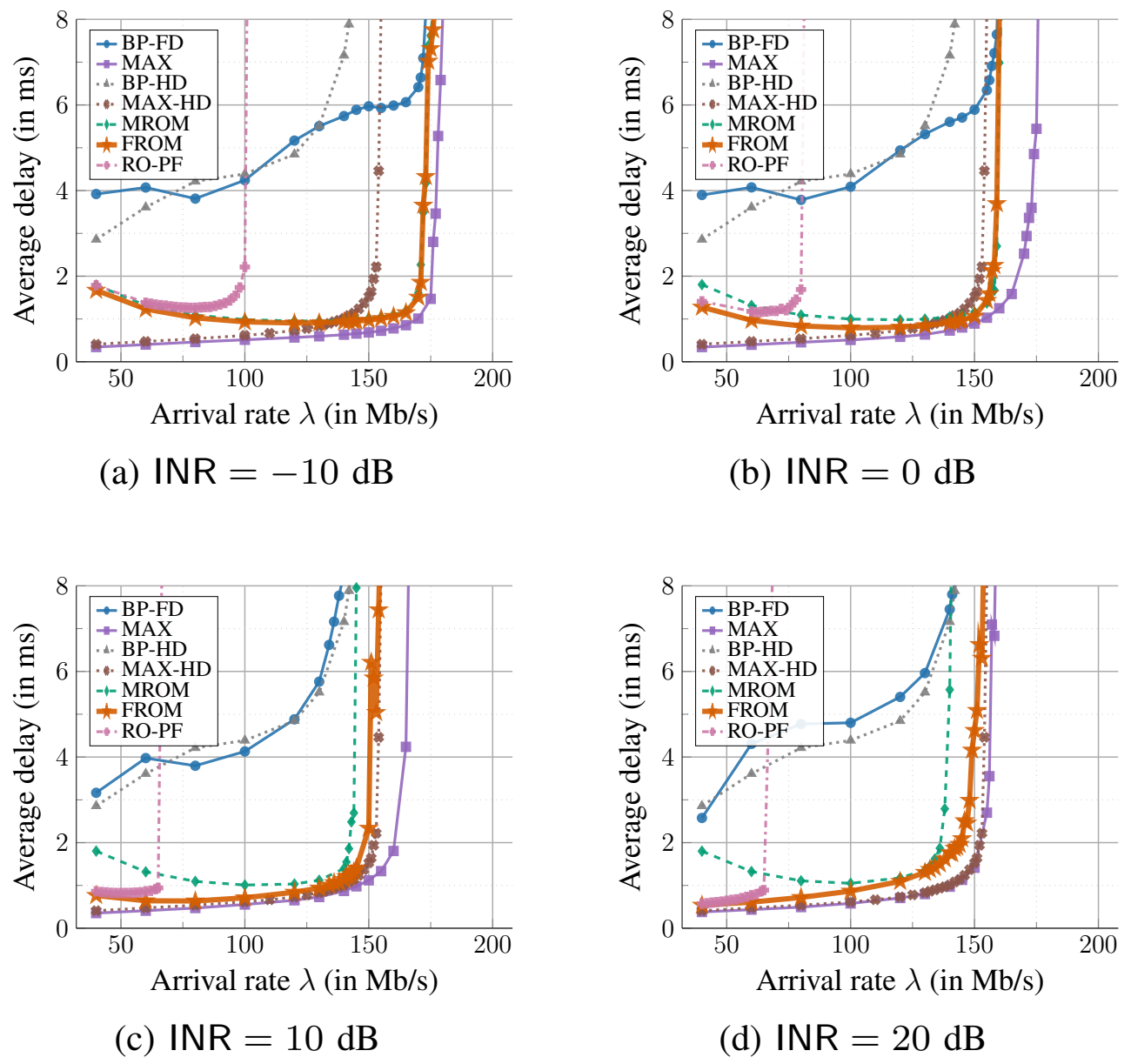

\centering
\subfloat[$\inr=-10$ dB]{%
    \resizebox{0.48\columnwidth}{!}{%
        \input{fig/lambda_sweep/lambda_inr-10.tex}%
    }%
    \label{fig/lambda_inr-10}%
}\hfill
\subfloat[$\inr=0$ dB]{%
    \resizebox{0.48\columnwidth}{!}{%
        \input{fig/lambda_sweep/lambda_inr0.tex}%
    }%
    \label{fig/lambda_inr0}%
}

\par\medskip

\subfloat[$\inr=10$ dB]{%
    \resizebox{0.48\columnwidth}{!}{%
        \input{fig/lambda_sweep/lambda_inr10.tex}%
    }%
    \label{fig/lambda_inr10}%
}\hfill
\subfloat[$\inr=20$ dB]{%
    \resizebox{0.48\columnwidth}{!}{%
        \input{fig/lambda_sweep/lambda_inr20.tex}%
    }%
    \label{fig/lambda_inr20}%
}

\caption{Average delay versus arrival rate $\lambda$ at different $\inr$ levels.}
\label{fig/lambda_sweep}
\end{figure}

Let us first consider \figref{fig/lambda_sweep}, depicting the average delay against the arrival rate $\lambda$, at four self-interference levels, $\inr \in \{-10, 0, +10, +20\}\ \mathrm{dB}$.
Both BP-HD and MAX-HD behave identically across $\inr$ levels because they prohibit simultaneous transmission and reception at each IAB and avoid self-interference.
Under HD, receiving backhaul traffic and forwarding traffic to UEs or downstream IABs require separate slots, limiting the service available as the load increases.
MAX can perform these operations simultaneously with little rate loss at low $\inr$, supporting higher arrival rates before delay grows sharply.
This advantage narrows as stronger self-interference reduces the rates achievable during simultaneous transmission and reception.
The initial decrease of reduced overhead policies in delay with arrival rate is related to the backhaul queue-length requirements.
A backhaul can be activated only after sufficient traffic has accumulated in its queue.
At low arrival rates, waiting to meet this requirement can contribute substantially to delay.
On the other hand, as the arrival rate increases, the required packets accumulate more quickly, reducing this waiting time.
MROM bases this requirement on the $\inr$-independent rate \(R_b^+\), whereas FROM uses \(R_b^-\) and can therefore shorten the waiting time when stronger self-interference lowers \(R_b^-\). 
Without this requirement, overly frequent backhaul activation at high self-interference may leave receiving children with insufficient transmission opportunities and make their queues unstable.
A similar motivation for preserving downstream transmission opportunities underlies the backhaul queue-length requirement in the half-duplex IAB setting of \cite{gopalam2022distributed}.
By allowing backhaul transmissions with less accumulated traffic, FROM can deliver packets to downstream IABs sooner. 
These packets can then be served by UE links earlier, helping explain the lower delay of FROM compared with MROM.
RO-PF applies fairness to individual links without giving a backhaul additional weight for carrying the combined traffic of multiple downstream users.
This can favor lightly loaded links over a congested backhaul, whereas FROM increases the backhaul's priority as its queue grows and achieves lower delay even when $\lambda$ is high.

\begin{figure}[t]
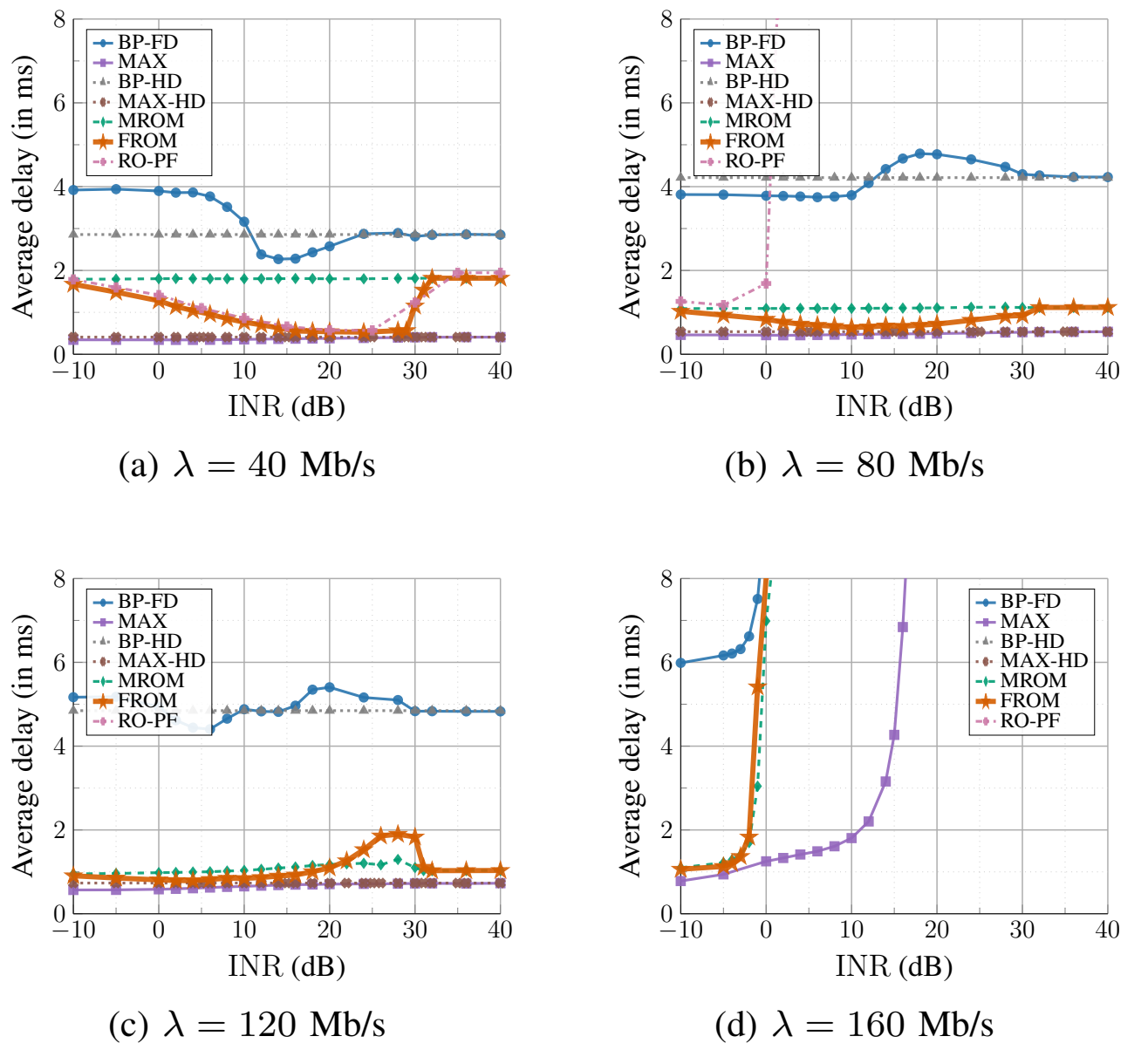

\centering
\subfloat[$\lambda=40$ Mb/s]{%
    \resizebox{0.48\columnwidth}{!}{%
        \input{fig/inr_sweep/inr_lam40.tex}%
    }%
    \label{fig:inr_lam40}%
}\hfill
\subfloat[$\lambda=80$ Mb/s]{%
    \resizebox{0.48\columnwidth}{!}{%
\begingroup
\providecommand{\falmplotwidth}{\linewidth}
\providecommand{\falmplotheight}{0.80\linewidth}
\pgfdeclareplotmark{falmstar}{%
\pgfpathmoveto{\pgfqpoint{0pt}{\pgfplotmarksize}}
\pgfpathlineto{\pgfqpoint{-.224514\pgfplotmarksize}{.309017\pgfplotmarksize}}
\pgfpathlineto{\pgfqpoint{-.951057\pgfplotmarksize}{.309017\pgfplotmarksize}}
\pgfpathlineto{\pgfqpoint{-.363271\pgfplotmarksize}{-.118034\pgfplotmarksize}}
\pgfpathlineto{\pgfqpoint{-.587785\pgfplotmarksize}{-.809017\pgfplotmarksize}}
\pgfpathlineto{\pgfqpoint{0pt}{-.381966\pgfplotmarksize}}
\pgfpathlineto{\pgfqpoint{.587785\pgfplotmarksize}{-.809017\pgfplotmarksize}}
\pgfpathlineto{\pgfqpoint{.363271\pgfplotmarksize}{-.118034\pgfplotmarksize}}
\pgfpathlineto{\pgfqpoint{.951057\pgfplotmarksize}{.309017\pgfplotmarksize}}
\pgfpathlineto{\pgfqpoint{.224514\pgfplotmarksize}{.309017\pgfplotmarksize}}
\pgfpathclose\pgfusepathqfillstroke}
\pgfdeclareplotmark{falmx}{%
\pgfpathmoveto{\pgfqpoint{-.5\pgfplotmarksize}{-1\pgfplotmarksize}}
\pgfpathlineto{\pgfqpoint{0pt}{-.5\pgfplotmarksize}}
\pgfpathlineto{\pgfqpoint{.5\pgfplotmarksize}{-1\pgfplotmarksize}}
\pgfpathlineto{\pgfqpoint{1\pgfplotmarksize}{-.5\pgfplotmarksize}}
\pgfpathlineto{\pgfqpoint{.5\pgfplotmarksize}{0pt}}
\pgfpathlineto{\pgfqpoint{1\pgfplotmarksize}{.5\pgfplotmarksize}}
\pgfpathlineto{\pgfqpoint{.5\pgfplotmarksize}{1\pgfplotmarksize}}
\pgfpathlineto{\pgfqpoint{0pt}{.5\pgfplotmarksize}}
\pgfpathlineto{\pgfqpoint{-.5\pgfplotmarksize}{1\pgfplotmarksize}}
\pgfpathlineto{\pgfqpoint{-1\pgfplotmarksize}{.5\pgfplotmarksize}}
\pgfpathlineto{\pgfqpoint{-.5\pgfplotmarksize}{0pt}}
\pgfpathlineto{\pgfqpoint{-1\pgfplotmarksize}{-.5\pgfplotmarksize}}
\pgfpathclose\pgfusepathqfillstroke}
\pgfdeclareplotmark{falmplus}{%
\pgfpathmoveto{\pgfqpoint{-.3333333333333333\pgfplotmarksize}{-1\pgfplotmarksize}}
\pgfpathlineto{\pgfqpoint{.3333333333333333\pgfplotmarksize}{-1\pgfplotmarksize}}
\pgfpathlineto{\pgfqpoint{.3333333333333333\pgfplotmarksize}{-.3333333333333333\pgfplotmarksize}}
\pgfpathlineto{\pgfqpoint{1\pgfplotmarksize}{-.3333333333333333\pgfplotmarksize}}
\pgfpathlineto{\pgfqpoint{1\pgfplotmarksize}{.3333333333333333\pgfplotmarksize}}
\pgfpathlineto{\pgfqpoint{.3333333333333333\pgfplotmarksize}{.3333333333333333\pgfplotmarksize}}
\pgfpathlineto{\pgfqpoint{.3333333333333333\pgfplotmarksize}{1\pgfplotmarksize}}
\pgfpathlineto{\pgfqpoint{-.3333333333333333\pgfplotmarksize}{1\pgfplotmarksize}}
\pgfpathlineto{\pgfqpoint{-.3333333333333333\pgfplotmarksize}{.3333333333333333\pgfplotmarksize}}
\pgfpathlineto{\pgfqpoint{-1\pgfplotmarksize}{.3333333333333333\pgfplotmarksize}}
\pgfpathlineto{\pgfqpoint{-1\pgfplotmarksize}{-.3333333333333333\pgfplotmarksize}}
\pgfpathlineto{\pgfqpoint{-.3333333333333333\pgfplotmarksize}{-.3333333333333333\pgfplotmarksize}}
\pgfpathclose\pgfusepathqfillstroke}
\begin{tikzpicture}
\begin{axis}[%
width=0.97\falmplotwidth,
height=\falmplotheight,
xmin=-10,
xmax=40,
ymin=0,
ymax=8,
font=\footnotesize,
axis background/.style={fill=white},
unbounded coords=jump,
filter discard warning=false,
scaled ticks=false,
xlabel={$\mathrm{INR}$ (dB)},
xlabel style={font={\fontsize{14}{16}\selectfont}},
ylabel={Average delay (in ms)},
ylabel style={font={\fontsize{14}{16}\selectfont}},
axis x line*=bottom,
axis y line*=left,
axis on top=false,
xtick={-10,0,10,20,30,40},
xmajorgrids=true,
minor xtick={-5,5,15,25,35},
xminorgrids=true,
ytick={0,2,4,6,8},
ymajorgrids=true,
minor ytick={1,3,5,7},
yminorgrids=true,
tick label style={font={\fontsize{12}{14}\selectfont}},
major grid style={solid,draw={rgb,1:red,0.680000;green,0.680000;blue,0.680000},line width=0.59999999999999998pt},
minor grid style={dotted,draw={rgb,1:red,0.820000;green,0.820000;blue,0.820000},line width=0.45000000000000001pt},
clip=true,
clip mode=individual,
restrict y to domain*={-0.80000000000000004:8.8000000000000007},
legend style={font={\fontsize{10}{11}\selectfont},at={(0.03,0.97)},anchor=north west,
  inner sep=1.8pt,row sep=0pt,nodes={inner xsep=1pt,inner ysep=0.5pt},
  legend cell align=left,draw=black,draw opacity=1,fill=white,fill opacity=0.9,text opacity=1},
legend image post style={mark indices={2},xscale=0.55,mark size=0.75\pgfplotmarksize},
]
\addplot [color={rgb,1:red,0.152941;green,0.454902;blue,0.682353}, line width=1.2800000000000002pt, solid, opacity=0.90000000000000002, mark=*, mark size=1.875pt, mark options={solid, fill={rgb,1:red,0.152941;green,0.454902;blue,0.682353}, draw={rgb,1:red,0.152941;green,0.454902;blue,0.682353}, line width=1pt}]
table[row sep=crcr]{%
-10	3.8115999999999999\\
-5	3.80829\\
0	3.7822399999999998\\
2	3.7778100000000001\\
4	3.7656299999999998\\
6	3.7475399999999999\\
8	3.7616700000000001\\
10	3.7958500000000002\\
12	4.0853700000000002\\
14	4.41927\\
16	4.6702599999999999\\
18	4.7888000000000002\\
20	4.7722100000000003\\
24	4.6511899999999997\\
28	4.4719199999999999\\
30	4.2964900000000004\\
32	4.2686799999999998\\
36	4.2317600000000004\\
40	4.2287499999999998\\
};
\addlegendentry{BP-FD}
\addplot [color={rgb,1:red,0.580392;green,0.403922;blue,0.741176}, line width=1.2800000000000002pt, solid, opacity=0.90000000000000002, mark=square*, mark size=1.875pt, mark options={solid, fill={rgb,1:red,0.580392;green,0.403922;blue,0.741176}, draw={rgb,1:red,0.580392;green,0.403922;blue,0.741176}, line width=1pt}]
table[row sep=crcr]{%
-10	0.46189999999999998\\
-5	0.45973000000000003\\
0	0.45567999999999997\\
2	0.45280999999999999\\
4	0.45382\\
6	0.46011000000000002\\
8	0.46704000000000001\\
10	0.47214\\
12	0.47865000000000002\\
14	0.48275000000000001\\
16	0.48508000000000001\\
18	0.48982999999999999\\
20	0.49408999999999997\\
24	0.50531000000000004\\
28	0.51907999999999999\\
30	0.52610000000000001\\
32	0.53103\\
36	0.53400000000000003\\
40	0.53686999999999996\\
};
\addlegendentry{MAX}
\addplot [color={rgb,1:red,0.498039;green,0.498039;blue,0.498039}, line width=1.2800000000000002pt, dotted, opacity=0.90000000000000002, mark=triangle*, mark size=1.875pt, mark options={solid, fill={rgb,1:red,0.498039;green,0.498039;blue,0.498039}, draw={rgb,1:red,0.498039;green,0.498039;blue,0.498039}, line width=1pt}]
table[row sep=crcr]{%
-10	4.2154199999999999\\
-5	4.2154199999999999\\
0	4.2154199999999999\\
2	4.2154199999999999\\
4	4.2154199999999999\\
6	4.2154199999999999\\
8	4.2154199999999999\\
10	4.2154199999999999\\
12	4.2154199999999999\\
14	4.2154199999999999\\
16	4.2154199999999999\\
18	4.2154199999999999\\
20	4.2154199999999999\\
24	4.2154199999999999\\
28	4.2154199999999999\\
30	4.2154199999999999\\
32	4.2154199999999999\\
36	4.2154199999999999\\
40	4.2154199999999999\\
};
\addlegendentry{BP-HD}
\addplot [color={rgb,1:red,0.549020;green,0.337255;blue,0.294118}, line width=1.2800000000000002pt, dotted, opacity=0.90000000000000002, mark=falmx, mark size=1.875pt, mark options={solid, fill={rgb,1:red,0.549020;green,0.337255;blue,0.294118}, draw={rgb,1:red,0.549020;green,0.337255;blue,0.294118}, line width=1pt}]
table[row sep=crcr]{%
-10	0.53869\\
-5	0.53869\\
0	0.53869\\
2	0.53869\\
4	0.53869\\
5	0.53869\\
6	0.53869\\
8	0.53869\\
10	0.53869\\
12	0.53869\\
14	0.53869\\
15	0.53869\\
16	0.53869\\
18	0.53869\\
20	0.53869\\
24	0.53869\\
25	0.53869\\
28	0.53869\\
30	0.53869\\
32	0.53869\\
35	0.53869\\
36	0.53869\\
40	0.53869\\
45	0.53869\\
50	0.53869\\
};
\addlegendentry{MAX-HD}
\addplot [color={rgb,1:red,0.000000;green,0.619608;blue,0.450980}, line width=1.2800000000000002pt, dashed, opacity=0.90000000000000002, mark=diamond*, mark size=1.875pt, mark options={solid, fill={rgb,1:red,0.000000;green,0.619608;blue,0.450980}, draw={rgb,1:red,0.000000;green,0.619608;blue,0.450980}, line width=1pt}]
table[row sep=crcr]{%
-10	1.0895600000000001\\
-5	1.0934900000000001\\
0	1.0929\\
2	1.09371\\
4	1.09476\\
6	1.0946899999999999\\
8	1.0958300000000001\\
10	1.0945499999999999\\
12	1.0974699999999999\\
14	1.09958\\
16	1.10185\\
18	1.10382\\
20	1.1077300000000001\\
24	1.12097\\
28	1.12382\\
30	1.1150800000000001\\
32	1.11598\\
36	1.1169899999999999\\
40	1.11778\\
};
\addlegendentry{MROM}
\addplot [color={rgb,1:red,0.835294;green,0.368627;blue,0.000000}, line width=2.4000000000000004pt, solid, opacity=0.90000000000000002, mark=falmstar, mark size=3.375pt, mark options={solid, fill={rgb,1:red,0.835294;green,0.368627;blue,0.000000}, draw={rgb,1:red,0.835294;green,0.368627;blue,0.000000}, line width=1pt}]
table[row sep=crcr]{%
-10	1.0259\\
-5	0.93335999999999997\\
0	0.83645000000000003\\
2	0.76622999999999997\\
4	0.71943999999999997\\
6	0.69089999999999996\\
8	0.66666999999999998\\
10	0.63671999999999995\\
12	0.65159\\
14	0.67632999999999999\\
16	0.66757999999999995\\
18	0.69213000000000002\\
20	0.71950000000000003\\
24	0.81225999999999998\\
28	0.90788999999999997\\
30	0.92442000000000002\\
32	1.11598\\
36	1.1169899999999999\\
40	1.11778\\
};
\addlegendentry{FROM}
\addplot [color={rgb,1:red,0.800000;green,0.474510;blue,0.654902}, line width=1.2800000000000002pt, dashdotted, opacity=0.90000000000000002, mark=falmplus, mark size=1.875pt, mark options={solid, fill={rgb,1:red,0.800000;green,0.474510;blue,0.654902}, draw={rgb,1:red,0.800000;green,0.474510;blue,0.654902}, line width=1pt}, mark indices={1,2,3,5}]
table[row sep=crcr]{%
-10	1.26166\\
-5	1.1731\\
0	1.68851\\
1.2686024467192158	8\\
5	26.56427\\
};
\addlegendentry{RO-PF}
\end{axis}
\end{tikzpicture}
\endgroup%
    }%
    \label{fig:inr_lam80}%
}

\par\medskip

\subfloat[$\lambda=120$ Mb/s]{%
    \resizebox{0.48\columnwidth}{!}{%
        \input{fig/inr_sweep/inr_lam120.tex}%
    }%
    \label{fig:inr_lam120}%
}\hfill
\subfloat[$\lambda=160$ Mb/s]{%
    \resizebox{0.48\columnwidth}{!}{%
        \input{fig/inr_sweep/inr_lam160.tex}%
    }%
    \label{fig:inr_lam160}%
}

\caption{Average delay versus $\inr$ at different arrival rates $\lambda$.}
\label{fig/inr_sweep}
\end{figure}

Fig. \ref{fig/inr_sweep} presents the average delay versus $\inr$ at fixed arrival rates.
The curves for RO-PF in Figs.~\ref{fig/inr_sweep}\subref{fig:inr_lam120}
and \ref{fig/inr_sweep}\subref{fig:inr_lam160}, and for BP-HD and MAX-HD in Fig. \ref{fig/inr_sweep}\subref{fig:inr_lam160}, are not plotted because they are unstable.
At $\lambda=160$ Mb/s, both BP-HD and MAX-HD are outside the displayed delay range because neither policy sustains this load in the experiment.
For \(\lambda=40,80,\) and \(120\) Mb/s, BP-FD approaches BP-HD at high self-$\inr$ as the behavior of BP-FD gets close to that of BP-HD. 
It can be observed that even though BP-FD has more flexibility than BP-HD, BP-HD exhibits lower delay than BP-FD.
This is because the back-pressure algorithm is designed to maximize throughput rather than minimize delay.
Also, under the DL-heavy traffic, sparsely arriving UL packets can retain small differential backlogs despite long waits, while BP-FD may favor concurrent DL transmissions that compete with UL reception and reduce its rate through self-interference. 
However, when self-interference is sufficiently weak and enough traffic is available for simultaneous reception and forwarding, BP-FD can reduce waiting by serving both in the same slot and sustain loads that BP-HD cannot accommodate, as observed at $\lambda=160$ Mb/s.
Stronger self-interference reduces backhaul rates during simultaneous reception and forwarding, while access downlink rates remain unaffected, potentially shifting RO-PF's priorities toward access links.
This can disadvantage backhaul transmissions precisely when they need more transmission opportunities to carry the same traffic, whereas FROM raises their priorities as queues accumulate.
For the reduced overhead policies, FROM and MROM coincide when self-interference is high and the maximum transmit-panel count becomes zero, yielding \(R_b^-=R_b^+\).
This transition increases both FROM's backhaul queue-length requirement and its backhaul scheduling weights.
For \(\lambda=40\) and \(80\) Mb/s, the additional accumulation time raises its delay. 
For \(\lambda=120\) Mb/s, however, the increased backhaul priority helps reduce accumulated queues, and the delay decreases instead. 
The different trends therefore reflect the competing effects of the queue-length requirement and backhaul prioritization.
The sharp delay growth at $\lambda=160$ Mb/s has a different explanation.
Unlike the lower arrival rates, this load cannot be sustained by either BP-HD or MAX-HD in the experiment.
Increasing self-interference reduces the rates achievable during simultaneous transmission and reception, eventually making the service insufficient to accommodate arrivals.

{\color{blue}}

\begin{figure}[t]
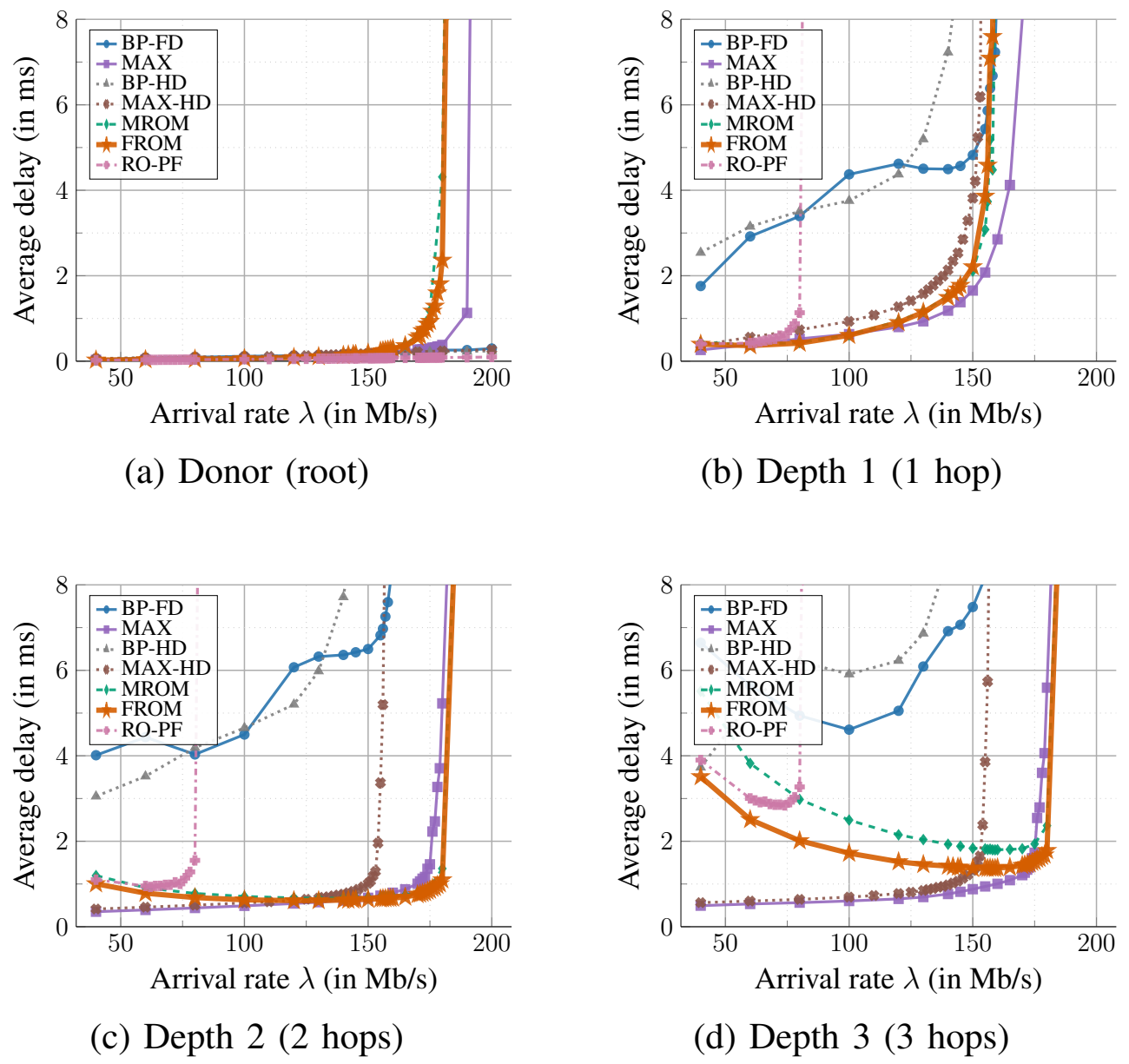

\centering
\subfloat[Donor (root)]{%
    \resizebox{0.48\columnwidth}{!}{%
        \input{fig/depth/depth0.tex}%
    }%
    \label{fig:depth0}%
}\hfill
\subfloat[Depth 1 (1 hop)]{%
    \resizebox{0.48\columnwidth}{!}{%
        \input{fig/depth/depth1.tex}%
    }%
    \label{fig:depth1}%
}

\par\medskip

\subfloat[Depth 2 (2 hops)]{%
    \resizebox{0.48\columnwidth}{!}{%
        \input{fig/depth/depth2.tex}%
    }%
    \label{fig:depth2}%
}\hfill
\subfloat[Depth 3 (3 hops)]{%
    \resizebox{0.48\columnwidth}{!}{%
        \input{fig/depth/depth3.tex}%
    }%
    \label{fig:depth3}%
}

\caption{Average delay versus $\lambda$ at different IAB tree depths for $\inr = 0$ dB.}
\label{fig:delay_by_depth}
\end{figure}

Fig. \ref{fig:delay_by_depth} compares the delay across different tree depths at $\inr=0$ dB.
Traffic associated with UEs attached to deeper IAB nodes traverses more wireless backhaul hops, so the delays incurred at successive hops accumulate along the route.
RO-PF maintains low delay for UEs attached to the root, but it becomes unstable at much lower arrival rates at depths 1–3.
While RO-PF can serve the UEs of the root well since their packets require no backhaul transmission, downstream packets can remain queued because its fairness criterion does not give backhaul links additional priority for carrying the traffic of multiple UEs.
The differences among the scheduling policies become particularly pronounced at depth 3, where FROM generally achieves a delay closer to the globally coordinated MAX policy than MROM.
This gap is consistent with the smaller backhaul queue-length requirements used by FROM.
In the considered topology, backhauls closer to the leaves aggregate traffic for fewer UEs, so accumulating sufficient traffic for backhaul activation can take longer at low arrival rates.
FROM permits downlink backhaul transmission with less accumulated traffic than MROM, which can reduce this waiting along multi-hop paths.
As the arrival rate increases, the queues fill more quickly, reducing the waiting associated with these requirements before congestion becomes dominant.
BP-HD and MAX-HD exhibit sharp delay growth at lower arrival rates at the non-donor depths, illustrating the benefit of simultaneous transmission and reception as the traffic load increases.

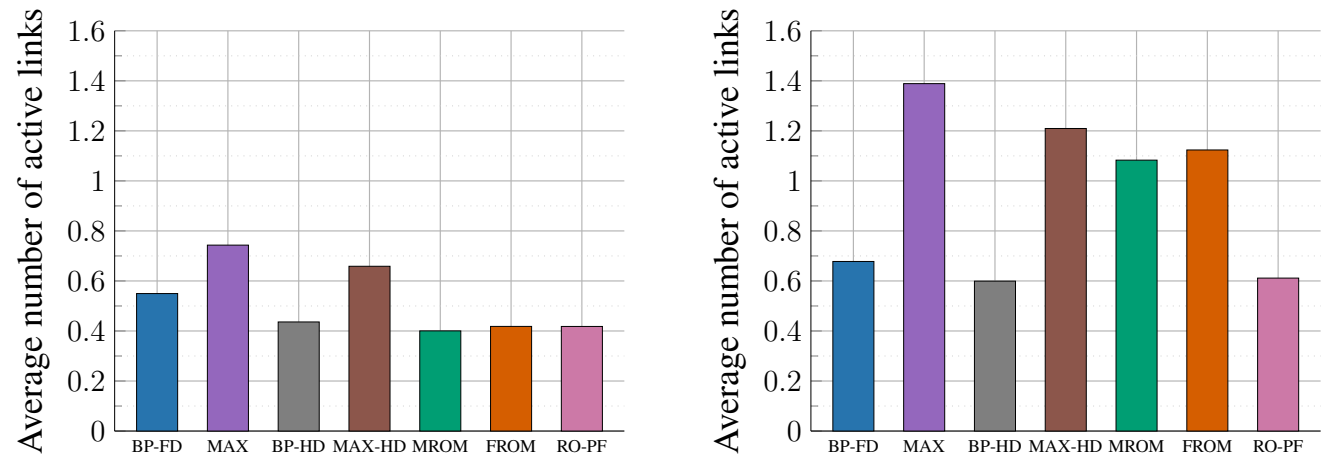
\begin{figure}[t]
\centering
\subfloat[$\lambda=40$ Mb/s]{%
    \resizebox{0.48\columnwidth}{!}{%
\begingroup
\providecommand{\falmplotwidth}{\linewidth}
\providecommand{\falmplotheight}{0.80\linewidth}
\begin{tikzpicture}
\begin{axis}[
width=0.97\falmplotwidth,height=\falmplotheight,
xmin=-0.6,xmax=6.6,ymin=0,ymax=1.6,
ytick distance=0.2,minor y tick num=1,scaled ticks=false,
font=\footnotesize,axis background/.style={fill=white},
axis x line*=bottom,axis y line*=left,
xtick={0,1,2,3,4,5,6},
xticklabels={{BP-FD},{MAX},{BP-HD},{MAX-HD},{MROM},{FROM},{RO-PF}},
xticklabel style={font=\scriptsize,align=center},
ylabel={Average number of active links},
ylabel style={font={\fontsize{14}{16}\selectfont}},
xmajorgrids,ymajorgrids,yminorgrids,
tick label style={font={\fontsize{12}{14}\selectfont}},
major grid style={draw=gray!60,line width=0.45pt},
minor grid style={draw=gray!35,dotted,line width=0.3pt},
clip=true]
\addplot+[ybar,solid,mark=none,bar shift=0pt,bar width=0.58,draw=black,line width=0.3pt,fill={rgb,255:red,39;green,116;blue,174}] coordinates {(0,0.549615151515152)};
\addplot+[ybar,solid,mark=none,bar shift=0pt,bar width=0.58,draw=black,line width=0.3pt,fill={rgb,255:red,148;green,103;blue,189}] coordinates {(1,0.743254545454546)};
\addplot+[ybar,solid,mark=none,bar shift=0pt,bar width=0.58,draw=black,line width=0.3pt,fill={rgb,255:red,127;green,127;blue,127}] coordinates {(2,0.43610303030303)};
\addplot+[ybar,solid,mark=none,bar shift=0pt,bar width=0.58,draw=black,line width=0.3pt,fill={rgb,255:red,140;green,86;blue,75}] coordinates {(3,0.658721212121212)};
\addplot+[ybar,solid,mark=none,bar shift=0pt,bar width=0.58,draw=black,line width=0.3pt,fill={rgb,255:red,0;green,158;blue,115}] coordinates {(4,0.400436363636364)};
\addplot+[ybar,solid,mark=none,bar shift=0pt,bar width=0.58,draw=black,line width=0.3pt,fill={rgb,255:red,213;green,94;blue,0}] coordinates {(5,0.4183)};
\addplot+[ybar,solid,mark=none,bar shift=0pt,bar width=0.58,draw=black,line width=0.3pt,fill={rgb,255:red,204;green,121;blue,167}] coordinates {(6,0.418127272727273)};
\end{axis}
\end{tikzpicture}
\endgroup%
    }%
    \label{fig:active_links_lam40}%
}\hfill
\subfloat[$\lambda=120$ Mb/s]{%
    \resizebox{0.48\columnwidth}{!}{%
\begingroup
\providecommand{\falmplotwidth}{\linewidth}
\providecommand{\falmplotheight}{0.80\linewidth}
\begin{tikzpicture}
\begin{axis}[
width=0.97\falmplotwidth,height=\falmplotheight,
xmin=-0.6,xmax=6.6,ymin=0,ymax=1.6,
ytick distance=0.2,minor y tick num=1,scaled ticks=false,
font=\footnotesize,axis background/.style={fill=white},
axis x line*=bottom,axis y line*=left,
xtick={0,1,2,3,4,5,6},
xticklabels={{BP-FD},{MAX},{BP-HD},{MAX-HD},{MROM},{FROM},{RO-PF}},
xticklabel style={font=\scriptsize,align=center},
tick label style={font={\fontsize{12}{14}\selectfont}},
ylabel={Average number of active links},
ylabel style={font={\fontsize{14}{16}\selectfont}},
xmajorgrids,ymajorgrids,yminorgrids,
major grid style={draw=gray!60,line width=0.45pt},
minor grid style={draw=gray!35,dotted,line width=0.3pt},
clip=true]
\addplot+[ybar,solid,mark=none,bar shift=0pt,bar width=0.58,draw=black,line width=0.3pt,fill={rgb,255:red,39;green,116;blue,174}] coordinates {(0,0.677921212121212)};
\addplot+[ybar,solid,mark=none,bar shift=0pt,bar width=0.58,draw=black,line width=0.3pt,fill={rgb,255:red,148;green,103;blue,189}] coordinates {(1,1.38883636363636)};
\addplot+[ybar,solid,mark=none,bar shift=0pt,bar width=0.58,draw=black,line width=0.3pt,fill={rgb,255:red,127;green,127;blue,127}] coordinates {(2,0.599306060606061)};
\addplot+[ybar,solid,mark=none,bar shift=0pt,bar width=0.58,draw=black,line width=0.3pt,fill={rgb,255:red,140;green,86;blue,75}] coordinates {(3,1.20953333333333)};
\addplot+[ybar,solid,mark=none,bar shift=0pt,bar width=0.58,draw=black,line width=0.3pt,fill={rgb,255:red,0;green,158;blue,115}] coordinates {(4,1.08301212121212)};
\addplot+[ybar,solid,mark=none,bar shift=0pt,bar width=0.58,draw=black,line width=0.3pt,fill={rgb,255:red,213;green,94;blue,0}] coordinates {(5,1.12367272727273)};
\addplot+[ybar,solid,mark=none,bar shift=0pt,bar width=0.58,draw=black,line width=0.3pt,fill={rgb,255:red,204;green,121;blue,167}] coordinates {(6,0.611451515151515)};
\end{axis}
\end{tikzpicture}
\endgroup%
    }%
    \label{fig:active_links_lam120}%
}
\caption{Average number of active links at $\inr =0$ dB.}
\label{fig:active_links}
\end{figure}

Fig. \ref{fig:active_links} depicts the average number of active links at $\inr =0$ dB for $\lambda=40$ and $120$ Mb/s. 
BP-FD and MAX activate more links than their HD counterparts because FD permits simultaneous transmission and reception at an IAB. 
These additional transmission opportunities can reduce packet waiting times and contribute to lower delay. 
FROM also exhibits a higher average number of active links than MROM. 
For the same queue state, using $R_b^-$ instead of $R_b^+$ lowers the weight of an outgoing backhaul link relative to other links. 
This can favor schedules that leave a child's incoming backhaul idle, allowing more transmissions at the child without the need to protect backhaul reception from self-interference.
The higher link activity observed under FROM is consistent with this scheduling preference and can contribute to its lower delay relative to MROM.
RO-PF activates approximately as many links as FROM at $\lambda=40$ Mb/s, but substantially fewer at $\lambda=120$ Mb/s.
Insufficient backhaul forwarding can leave packets queued at the root rather than supplying downstream IABs, reducing their link activity despite the growing network backlog.

Fig. \ref{fig:control_delay} compares the average delay against the control time $\delta$ for $\lambda=40$ and $120$ Mb/s at $\inr=0$ dB.
RO-PF is omitted from Fig.~\ref{fig:control_delay}\subref{fig:control_lam120} since the policy is unstable throughout all $\delta$.
Control information passes through successive levels of the IAB tree, so the delays incurred during these exchanges accumulate.
This accumulated overhead reduces the time available to serve traffic along multi-hop routes.
For $\lambda=40$ Mb/s, the remaining transmission time is sufficient to accommodate arrivals, 
allowing global policies to maintain lower delay as long as they are stable.
On the other hand, when $\delta$ reaches around $20~\mu$s, the maximum accumulated control delay of $6\delta$ under global policies occupies $120~\mu$s of the $125~\mu$s slot, leaving little time for packet transmission at downstream nodes.
Reduced-overhead policies retain more transmission time and remain stable up to substantially larger values of $\delta$, as the maximum accumulated control delay of the reduced-overhead policy is $4\delta$, and it can endure up to $T_s/4 \approx 31~\mu s$.
For $\lambda=120$ Mb/s, global policies become unstable at smaller values of $\delta$, whereas reduced overhead policies tolerate larger control delays.
At this higher arrival rate, the loss of transmission time due to control overhead has a greater impact on packet delay, especially for deeper nodes.
Consequently, reduced overhead policies can achieve lower packet delay than global policies and maintain low delay even when the latter become unstable.

\section{Conclusion} \label{sec:conclusion}
This paper presented distributed and reduced overhead scheduling algorithms for full-duplex IAB networks.
Self-interference makes link weights depend on other scheduling decisions, complicating the scheduling problem.
We addressed this dependence through dynamic programming, enabling efficient computation of globally optimal schedules.
We also proposed FROM, which reduces control overhead by relying only on information from neighboring nodes.
We established a provable lower bound on its downlink stability region and showed that FROM matches the global stability region under specific channel and interference conditions.
Numerical results demonstrated that full-duplex operation can support a wider range of traffic loads than half-duplex operation.
They also showed that FROM can achieve delay performance and observed stability limits close to those of global scheduling, despite its limited coordination.
The current stability analysis, however, is restricted to downlink traffic.
Extending the analysis to joint downlink and uplink traffic and narrowing the remaining stability gap without substantially increasing control overhead are directions for future work.

\appendices
\section{Derivation of the Dynamic Program}
\label{app:dp-derivation}

The objective in \eqref{eq:mws} decomposes as
\begin{equation}
\label{eq:app-dp-link-partition}
\begin{aligned}
&\sum_{\ell\in\mathcal{L}(\mathcal{G}_n)}w_\ell(m_\ell)s_\ell
=\sum_{\ell\in\mathcal{L}_n\cap\mathcal{L}_{\mathrm{AC}}}w_\ell(m_\ell)s_\ell\\
&+\sum_{c\in\mathcal{C}_n}\Bigl[
    w_{b_c^{\mathrm{DL}}}(m_c)s_{b_c^{\mathrm{DL}}}
    +w_{b_c^{\mathrm{UL}}}(m_n)s_{b_c^{\mathrm{UL}}}+\sum_{\ell\in\mathcal{L}(\mathcal{G}_c)}w_\ell(m_\ell)s_\ell
\Bigr].
\end{aligned}
\end{equation}
Consider $\Phi_n^\theta(m)$ with inherited mode $\theta_n=\theta$.
For a given local schedule $\vs_n$ and $m$, the optimal contribution of child $c$ is $\nu_c^T$, $\nu_c^R(m)$, or $\nu_c^{\IDLE}$.
Thus, subtracting each child's idle value before adding its active value gives
\begin{equation}
\label{eq:app-dp-effective-weight}
\begin{aligned}
&\Phi_n^\theta(m)
=\sum_{c\in\mathcal{C}_n}\nu_c^{\IDLE}
+\max_{\vs_n}\Bigl\{
    \sum_{\ell\in\mathcal{L}_n\cap\mathcal{L}_{\mathrm{AC}}}w_\ell(m_\ell)s_\ell\\
&\quad+\sum_{c\in\mathcal{C}_n:\,\theta_c=T}(\nu_c^T-\nu_c^{\IDLE})
+\sum_{c\in\mathcal{C}_n:\,\theta_c=R}(\nu_c^R(m)-\nu_c^{\IDLE})
\Bigr\},
\end{aligned}
\end{equation}
where the maximum is over feasible local schedules with inherited mode $\theta_n=\theta$ and total TX count $m$ at node $n$.

Now, we will compute the maximum weight of each panel when each panel picks a transmitting or reception link, $\psi_{n,k}^T$ and $\psi_{n,k}^R(m)$, respectively.
If the maximizing link on panel $k$ is a UE access link, $\ell$, $w_\ell$ or $w_\ell(m)$ should be selected as a $\psi_{n,k}^T$ or $\psi_{n,k}^R(m)$ depending on whether the link is used for transmission or for reception.
When the maximizing link is a child backhaul link, $b_c^{\mathrm{DL}}$ or $b_c^{\mathrm{UL}}$, its candidate value is $\nu_c^T$ or $\nu_c^R(m)$, respectively, accounting for both the backhaul link and the optimal schedule within the child's subtree.
These cases correspond to $\tau_\ell$ and $\rho_\ell(m)$.
Even when the child backhaul is not selected, links within that subtree may remain active, so its idle-mode contribution must still be included.
The term $\phi_\ell$ represents this idle-mode contribution and is subtracted from the active-mode value to avoid double counting it in the baseline sum.
Because each panel selects at most one link, in $\psi_{n,k}^T$, $\max$ picks only one link and in $\psi_{n,k}^R(m)$, it has a similar form, but it also allows the panel to remain idle, as represented by $(\cdot)^+$.
\begin{equation}
\label{eq:app-dp-panel-values}
\begin{aligned}
\psi_{n,k}^T
&=B_{n,k}
+\max_{\ell\in\mathcal L_{n,k}^{\mathrm{TX}}
\cap\mathcal L(\mathcal G_n)}(\tau_{\ell}-\phi_{\ell}),\\
\psi_{n,k}^R(m)
&=B_{n,k}
+\bigl[
\max_{\ell\in\mathcal L_{n,k}^{\mathrm{RX}}
\cap\mathcal L(\mathcal G_n)}
(\rho_\ell(m)-\phi_\ell)
\bigr]^+.
\end{aligned}
\end{equation}

Next, we will obtain the node $n$'s optimal sums of weights, $\nu_n^T$, $\nu_n^R(m')$, and $\nu_n^{\IDLE}$, by combining $\psi_{n,k}^T$ and $\psi_{n,k}^R(m)$.
Before getting $\Phi_n^\theta(m)$, we will first divide panels into transmitting panels and reception/idle panels, which are denoted as $P$ and $\mathcal{P}_n'\setminus P$, respectively.
Note that $\mathcal{P}_n'$ is from Algorithm~\ref{alg:dp}: $\mathcal{P}_n\setminus\{P_n^1\}$ for $\theta\in\{T,R\}$ and $\mathcal{P}_n$ for $\theta=\IDLE$.
To use $P$ as an optimization variable, we reformulate the sum of weights as follows:
\[
\begin{aligned}
&\sum_{k\in P}\psi_{n,k}^T
+\sum_{k\in\mathcal{P}_n'\setminus P}\psi_{n,k}^R(m)\\
&\quad=\sum_{k\in\mathcal{P}_n'}\psi_{n,k}^R(m)
+\sum_{k\in P}[\psi_{n,k}^T-\psi_{n,k}^R(m)].
\end{aligned}
\]
If $\theta\in\{T,R\}$, all child backhaul links in $P_n^1$ should be idle.
Hence, for $\theta\in\{T,R\}$,
\begin{equation}
\label{eq:app-dp-fixed-count}
\begin{aligned}
\Phi_n^\theta(m)
&=\sum_{\substack{c\in\mathcal{C}_n:\,b_c^{\mathrm{DL}}\in\mathcal{L}_{n,P_n^1}}}\nu_c^{\IDLE}
+\sum_{k\in\mathcal{P}_n'}\psi_{n,k}^R(m)\\
&\quad+\max_{\substack{P\subseteq\mathcal{P}_n'\\|P|=m-\mathbf{1}\{\theta=R\}}}
\sum_{k\in P}[\psi_{n,k}^T-\psi_{n,k}^R(m)].
\end{aligned}
\end{equation}

For $\theta=\IDLE$, the children links in $P_n^1$ can be in other states, thus,
\begin{equation}
\label{eq:app-dp-fixed-count-idle}
\begin{aligned}
\Phi_n^{\IDLE}(m)
&=\sum_{k\in\mathcal{P}_n}\psi_{n,k}^R(m)+\max_{\substack{P\subseteq\mathcal{P}_n\\|P|=m}}
\sum_{k\in P}\{\psi_{n,k}^T-\psi_{n,k}^R(m)\}.
\end{aligned}
\end{equation}

Adding the parent-backhaul weight gives
\begin{equation}
\label{eq:app-dp-boundary-values}
\begin{aligned}
\nu_n^T
&=\max_m[w_{b_n^{\mathrm{DL}}}(m)+\Phi_n^T(m)],\\
\nu_n^R(m')
&=w_{b_n^{\mathrm{UL}}}(m')+\max_m\Phi_n^R(m),\\
\nu_n^{\IDLE}
&=\max_m\Phi_n^{\IDLE}(m).
\end{aligned}
\end{equation}

\section{Proof of Theorem~\ref{thm:reduced-overhead-policy-region}}
\label{app:reduced-overhead-policy-region}

\subsection{Proof of 1)}
We prove the first statement by contraposition.
Suppose that a stationary reduced overhead policy stabilizes the network under arrival-rate vector $\vlambda$.
By the definition of $\overline{\mathcal C}_{\mathrm{ro}}$ and the independence of the current channel state from the queues, $\mathbb E[\vc(t)\mid\boldsymbol Q(t)]\in\overline{\mathcal C}_{\mathrm{ro}}$.
As $\overline{\mathcal C}_{\mathrm{ro}}$ is defined as a convex hull, averaging over the random choices, queue states, and slots keeps the mean service vector in $\overline{\mathcal C}_{\mathrm{ro}}$ as well.
Its service vector $\vc(t)$ therefore satisfies
\begin{equation}
\label{eq:reduced-overhead-policy-mean-service}
\frac1T\sum_{t=0}^{T-1}\mathbb E[\vc(t)]
\in\overline{\mathcal C}_{\mathrm{ro}}.
\end{equation}

Under stability, bounded expected queue increments imply $\mathbb E[q_n^f(T)]/T\to0$.
Along the root-to-$n$ route $n_0=r,\ldots,n_h=n$, as $T \rightarrow \infty$,
\begin{equation}
\label{eq:reduced-overhead-policy-route-telescoping}
\begin{aligned}
\frac1T\sum_{t=0}^{T-1}\mathbb E[d_n^f(t)]
&=\lambda_f-\frac1T\sum_{i=0}^h
\mathbb E[q_{n_i}^f(T)-q_{n_i}^f(0)]\\
&\longrightarrow\lambda_f.
\end{aligned}
\end{equation}
By compactness of $\overline{\mathcal C}_{\mathrm{ro}}$, choose a subsequence $T_i$ along which \eqref{eq:reduced-overhead-policy-mean-service} converges to $\overline{\vc}\in\overline{\mathcal C}_{\mathrm{ro}}$.
Summing \eqref{eq:reduced-overhead-policy-route-telescoping} over $f\in\mathcal F_\ell$ and bounding departures by offered service gives
\begin{equation}
\label{eq:reduced-overhead-policy-necessary-region}
\lambda_\ell
\le\lim_{i\to\infty}\frac1{T_i}\sum_{t=0}^{T_i-1}
\mathbb E[c_\ell(t)]
=\overline c_\ell.
\end{equation}
Thus $\vlambda\in\Lambda_{\mathrm{ro}}$.

\subsection{Proof of 2)}

Suppose $\vlambda+\varepsilon\boldsymbol1\in\Lambda_{\mathrm{ro}}$ for some $\varepsilon>0$.
By the region definition and Carath\'eodory's theorem, there are $K\le|\mathcal L_{\mathrm{DL}}|+1$ deterministic reduced overhead rules $d_j\in\mathfrak D_{\mathrm{ro}}$ and coefficients $\alpha_j\ge0$, $\sum_{j=1}^K\alpha_j=1$, such that
\begin{equation}
\label{eq:reduced-overhead-policy-initial-slack}
\begin{aligned}
\overline c_\ell
&=\sum_{j=1}^K\alpha_j\mathbb E[c_\ell^{d_j}(t)]\ge\lambda_\ell+\varepsilon|\mathcal F_\ell|
\ge\lambda_\ell+\varepsilon.
\end{aligned}
\end{equation}

Following the standard stationary randomized construction in \cite[Lemma~8]{neely2003dynamic}, let $\Pr^\alpha$ denote the distribution obtained by choosing $d_j$ with probability $\alpha_j$.
Construct policy $\pi$ in root-to-leaf order: starting with $\theta_r=\IDLE$, each IAB selects its schedule using the probabilities under $\Pr^\alpha$, given its parent's backhaul decision and local SNRs, with independent random draws at each IAB.
By construction, $\pi$ has the same conditional selection probabilities as those under $\Pr^\alpha$.
As each parents' decision is independent of its children's information,
\begin{equation}
\label{eq:reduced-overhead-policy-stationary-service}
\mathbb E^\pi[c_\ell(t)]
=\sum_{j=1}^K\alpha_j\mathbb E[c_\ell^{d_j}(t)]
=\overline c_\ell.
\end{equation}

On each selected link, packet departure is the smaller of the queue length and the link rate.
\begin{equation}
\label{eq:reduced-overhead-policy-work-conserving-service}
\sum_{f\in\mathcal F_\ell}d_n^f(t)
=\min\{Q_\ell(t),c_\ell(t)\}.
\end{equation}

Let the superscript $x$ index systems with initial total backlog $\sum_\ell Q_\ell^x(0)=x$.
For any sequence $x_k\to\infty$, the standard fluid-limit construction \cite{dai1995positive,Dai2000throughput} gives, almost surely, a subsequence $x_{k_j}$ along which, for every node $n$ and $\ell\in\mathcal L_n^{\mathrm{TX}}$,
\begin{align}
\frac{Q_\ell^{x_{k_j}}(\lfloor x_{k_j}\tau\rfloor)}{x_{k_j}}
&\longrightarrow\bar Q_\ell(\tau),
\label{eq:FROM-fluid-queue-limit}\displaybreak[1]\\
\frac1{x_{k_j}}\sum_{t=0}^{\lfloor x_{k_j}\tau\rfloor-1}
\sum_{f\in\mathcal F_\ell}a_n^{f,x_{k_j}}(t)
&\longrightarrow\int_0^\tau\bar a_\ell(u)\,du,
\label{eq:FROM-fluid-arrival-limit}\displaybreak[1]\\
\frac1{x_{k_j}}\sum_{t=0}^{\lfloor x_{k_j}\tau\rfloor-1}
\sum_{f\in\mathcal F_\ell}d_n^{f,x_{k_j}}(t)
&\longrightarrow\int_0^\tau\bar r_\ell(u)\,du,
\label{eq:FROM-fluid-departure-limit}
\end{align}
uniformly on compact time intervals.
The vector $\bar{\boldsymbol Q}(\tau)$ collects these aggregate fluid queues.
Thus, $\bar a_\ell$ and $\bar r_\ell$ are the derivatives of the cumulative arrival and departure limits.
The queue equation gives
\begin{equation}
    \dot{\bar Q}_\ell(\tau)=\bar a_\ell(\tau)-\bar r_\ell(\tau).
\label{eq:FROM-fluid-balance}
\end{equation}

Now we will show that in the fluid-limit model, as long as $\bar Q(\tau) > 0$, departure from the queue is same to $\bar c_\ell$.
Assume that $\bar Q_\ell(\tau)>0$ and since $c_\ell (t)$ is finite, for sufficiently large $x_{k_j}$,
\[
Q_\ell^{x_{k_j}}(\lfloor x_{k_j}\tau\rfloor)
\sim x_{k_j}\bar Q_\ell(\tau)\longrightarrow\infty.
\]
By \cite[Lemma~5(i)]{stolyar2004maxweight}, this implies that $Q_\ell^{x_{k_j}}(t)\ge c_\ell(t)$ throughout the slots corresponding to a sufficiently short interval after $\tau$, for all sufficiently large $j$.

At the root, $\bar a_\ell=\lambda_\ell$. 
In the other nodes, after all ancestor queues remain zero, the same holds at each subsequent depth because their arrivals and departures agree.
Thus from \eqref{eq:reduced-overhead-policy-initial-slack}, every positive fluid queue at that depth satisfies
\begin{equation}
\label{eq:reduced-overhead-policy-draining-drift}
\dot{\bar Q}_\ell(\tau)
\le\lambda_\ell-\overline c_\ell
\le-\varepsilon.
\end{equation}
Forwarding packets between IABs removes them from the transmitting IAB's queue and adds them to the receiving IAB's queue, leaving the sum of their backlogs unchanged.
Thus, with the normalization $\sum_\ell\bar Q_\ell(0)=1$, the total backlog can increase only through new traffic arriving at the root, giving
\begin{equation}
\label{eq:FROM-total-fluid-bound}
\sum_{\ell\in\mathcal L_{\mathrm{DL}}}\bar Q_\ell(\tau)
\le 1+\tau\sum_{f\in\mathcal F_{\mathrm{DL}}}\lambda_f.
\end{equation}
Whenever $\bar Q_\ell(\tau)>0$, its departure rate exceeds its arrival rate by at least $\varepsilon$, so $\dot{\bar Q}_\ell(\tau)\le-\varepsilon$.
Since the initial total backlog is one, all root queues become zero by time $t_1=1/\varepsilon$ and remain zero.

Now suppose that all queues transmitted at depths smaller than $i$ remain zero after time $t_i$.
Then, their arrival rates are same to the departure rates, so each queue at depth $i$ receives traffic at rate $\lambda_\ell$.
By \eqref{eq:reduced-overhead-policy-draining-drift}, each positive queue at this depth therefore decreases at rate at least $\varepsilon$.
Moreover, \eqref{eq:FROM-total-fluid-bound} gives
\[
\bar Q_\ell(t_i)
\le 1+t_i\sum_{f\in\mathcal F_{\mathrm{DL}}}\lambda_f.
\]
Hence, all queues at depth $i$ become zero and remain zero by time
\[
t_{i+1}
=t_i+\frac{1+t_i\sum_{f\in\mathcal F_{\mathrm{DL}}}\lambda_f}
{\varepsilon}.
\]
Starting with $t_0=0$, this argument applies successively to every depth.
Since the tree has finite depth, all fluid limits drain within a common finite time.
Stability then follows from \cite{dai1995stability} and the policy $\pi$ can stabilize the network.

\section{Proof of Proposition~\ref{prop:reduced overhead-global-region}}
\label{app:local-global-region}

Since $\Lambda_{\mathrm{ro}}\subseteq\Lambda$ is trivial, we will prove $\frac{1}{2}\Lambda \subseteq \Lambda_{\mathrm{ro}}$.
First, choose $\vlambda\in\Lambda$.
By the definition of $\Lambda$, there exists a randomized schedule $\vs^{\mathrm g}(t)$, selected according to the current channel state, whose mean service vector, $\overline{\vc}$ satisfies
\begin{equation}
\label{eq:parity-global-service}
\begin{aligned}
\overline c_\ell
&=\mathbb E\!\left[
R_\ell\!\left(m_\ell(\vs^{\mathrm g}(t),t),t\right)
s_\ell^{\mathrm g}(t)
\right]\ge\lambda_\ell,
\end{aligned}
\end{equation}

For \(p\in\{0,1\}\), let \(\mathcal B_p\) contain the IABs whose depth satisfies depth \(\bmod\ 2=p\), with the root at depth zero.
Let $\vs_{n,\mathrm{out}}$ collect the outgoing activations of IAB $n$, and let $\snr_n(t)$ denote its channel states.
Construct a randomized reduced overhead rule under which only IABs in $\mathcal B_p$ may transmit.
Each such IAB independently samples the joint activation of its outgoing links according to
\begin{equation}
\label{eq:parity-local-marginal}
\begin{aligned}
&\Pr\!\left\{
\vs_{n,\mathrm{out}}^{(p)}(t)=\boldsymbol u
\mid\snr_n(t)
\right\}=
\Pr\!\left\{
\vs_{n,\mathrm{out}}^{\mathrm g}(t)=\boldsymbol u
\mid\snr_n(t)
\right\},
\end{aligned}
\end{equation}
whereas IABs outside $\mathcal B_p$ turn off all of their outgoing activations; thus, it operates in HD mode, and there is no self-interference.
Note that the IAB $n$ determines its scheduling under the sampled policy, it does not require the information of its children, therefore the policy follows a reduced overhead scheduling constraint.

Let $\overline{\vc}^{(p)}\in\overline{\mathcal C}_{\mathrm{ro}}$ denote the resulting mean offered-service vector.
A downlink transmitted by an IAB in $\mathcal B_p$ provides zero service when only IABs in $\mathcal B_{1-p}$ transmit.
When only IABs in $\mathcal B_p$ transmit, its receiver does not transmit, so its rate is $R_\ell(0,t)$.
By \eqref{eq:parity-local-marginal}, its activation probability matches the global policy for the same channel states, while removing self-interference cannot decrease its rate.
Therefore, summing the average service of the same link under these two policies gives, for every $\ell\in\mathcal L_{\mathrm{DL}}$,
\begin{equation}
\label{eq:parity-service-domination}
\begin{aligned}
\overline c_\ell^{(0)}
+\overline c_\ell^{(1)}
&=\mathbb E\!\left[
R_\ell(0,t)s_\ell^{\mathrm g}(t)
\right]\\
&\ge\mathbb E\!\left[
R_\ell\!\left(m_\ell(\vs^{\mathrm g}(t),t),t\right)
s_\ell^{\mathrm g}(t)
\right]\\
&=\overline c_\ell
\ge\lambda_\ell.
\end{aligned}
\end{equation}

By convexity of $\overline{\mathcal C}_{\mathrm{ro}}$,
\begin{equation}
\label{eq:parity-service-average}
\frac{\overline{\vc}^{(0)}
+\overline{\vc}^{(1)}}{2}
\in\overline{\mathcal C}_{\mathrm{ro}}.
\end{equation}
The local service vector provides at least $\lambda_\ell/2$ on every downlink, so $\vlambda/2\in\Lambda_{\mathrm{ro}}$ by definition.
Since this holds for every $\vlambda\in\Lambda$, we obtain $\frac12\Lambda\subseteq\Lambda_{\mathrm{ro}}$.
\hfill$\square$

\section{Proof of Theorem~\ref{thm:FROM-stability}}
\label{app:FROM-stability}

Pick $\vlambda\in\operatorname{int}(\beta\Lambda_{\mathrm{ro}})$ and choose $\varepsilon>0$ such that $\vlambda/\beta+\varepsilon\boldsymbol1\in\Lambda_{\mathrm{ro}}$.
Choose $\overline{\vc}$ and the reduced overhead rules as in \eqref{eq:reduced-overhead-policy-initial-slack}, now for the load $\vlambda/\beta$.
By \eqref{eq:FROM-reduced-overhead-arrival-region} and $|\mathcal F_\ell|\ge1$, they satisfy
\begin{equation}
\label{eq:FROM-reference-slack}
\beta\overline c_\ell
\ge\lambda_\ell+\beta\varepsilon|\mathcal F_\ell|
\ge\lambda_\ell+\delta,
\qquad \delta\eqdef\beta\varepsilon>0.
\end{equation}
For every non-root IAB $n$, define
\begin{equation}
\label{eq:FROM-reference-mode}
\rho_n\eqdef\sum_{j=1}^K\alpha_j
\mathbb E\!\left[s_{b_n^{\mathrm{DL}}}^{d_j}(t)\right].
\end{equation}
where $s_{b_n^{\mathrm{DL}}}^{d_j}(t) \in \{0,1\}$ is the binary variable of the downlink backhaul from $\mathrm{pr}(n)$ to $n$ under rule $d_j$ in slot $t$.

The maximum backhaul rate, $R_{b_n^{\mathrm{DL}}}^{+}$, gives
\begin{equation}
\label{eq:FROM-reference-upper-bound}
\overline c_{b_n^{\mathrm{DL}}}\le R_{b_n^{\mathrm{DL}}}^{+}\rho_n.
\end{equation}

For FROM, use \eqref{eq:FROM-fluid-queue-limit}--\eqref{eq:FROM-fluid-departure-limit}, with the subsequence and limits taken for the FROM.
Along the same subsequence, define the fluid activation rate $\bar s_\ell$ by
\begin{equation}
\label{eq:FROM-fluid-activation-limit}
\frac1{x_{k_j}}\sum_{t=0}^{\lfloor x_{k_j}\tau\rfloor-1}s_\ell^{x_{k_j}}(t)
\longrightarrow\int_0^\tau\bar s_\ell(u)\,du,
\end{equation}
uniformly on compact time intervals.
A regular time is one at which the queue and cumulative-limit derivatives exist, which holds almost everywhere with \eqref{eq:FROM-fluid-balance} and \eqref{eq:FROM-total-fluid-bound}.

For a fixed aggregate-queue vector $\boldsymbol Q$, write the local objective without the parent backhaul as
\begin{equation}
\label{eq:FROM-outgoing-objective}
\begin{aligned}
W_n(\vs_n,t;\boldsymbol Q)\eqdef{}
&\sum_{\ell\in\mathcal L_n^{\mathrm{TX}}\cap\mathcal L_{\mathrm{AC}}}
Q_\ell R_\ell(t)s_\ell+\sum_{c\in\mathcal C_n}Q_{b_c^{\mathrm{DL}}}
R_{b_c^{\mathrm{DL}}}^{-}s_{b_c^{\mathrm{DL}}}.
\end{aligned}
\end{equation}
Write $W_n(\vs_n,t)=W_n(\vs_n,t;\boldsymbol Q(t))$ for simplicity.
Let $\widehat{\mathcal S}_n^\theta(t)$ be the downlink feasible set in \eqref{eq:FROM-feasible-set} with the queue-eligibility constraints removed.
Define
\begin{align}
M_n^{\IDLE}(\boldsymbol Q)
&\eqdef\mathbb E\!\left[
\max_{\vs_n\in\widehat{\mathcal S}_n^{\IDLE}(t)}
W_n(\vs_n,t;\boldsymbol Q)\right].
\label{eq:FROM-mode-maximum}\\
M_n^T(\boldsymbol Q)
&\eqdef\mathbb E\Bigl[
\max_{\vs_n\in\widehat{\mathcal S}_n^T(t)}W_n(\vs_n,t;\boldsymbol Q)\Bigr],
\label{eq:FROM-available-idle-mean}
\end{align}
The expectations are over the channel state distribution.
The corresponding values for fluid-limit $\bar Q_\ell(\tau)$ are obtained by replacing each $Q_\ell$ with $\bar Q_\ell(\tau)$.

\begin{lemma}
\label{lem:FROM-fluid-service}
For a non-root IAB $n$, FROM's departures satisfy
\begin{equation}
\label{eq:FROM-gated-departure}
\sum_{f\in\mathcal F_{b_n^{\mathrm{DL}}}}d_{\mathrm{pr}(n)}^f(t)
\ge R_{b_n^{\mathrm{DL}}}^{-}s_{b_n^{\mathrm{DL}}}(t).
\end{equation}
Moreover, $M_n^\IDLE(\boldsymbol Q)\ge M_n^T(\boldsymbol Q)$ for every nonnegative $\boldsymbol Q$, and at every regular fluid time,
\begin{align}
&\sum_{\ell\in\mathcal L_n^{\mathrm{TX}}}
\bar Q_\ell(\tau)\bar r_\ell(\tau)
\ge M_n^{\IDLE}(\bar{\boldsymbol Q}(\tau))
\nonumber\\
&\quad-\bar s_{b_n^{\mathrm{DL}}}(\tau)
\Bigl[M_n^\IDLE(\bar{\boldsymbol Q}(\tau))
-M_n^T(\bar{\boldsymbol Q}(\tau)) +\bar Q_{b_n^{\mathrm{DL}}}(\tau)R_{b_n^{\mathrm{DL}}}^{+}\Bigr].
\label{eq:FROM-fluid-weight-before-relaxation}
\end{align}
Since the root has no parent backhaul link,
\begin{equation}
\label{eq:FROM-root-own-maximum}
\sum_{\ell\in\mathcal L_r^{\mathrm{TX}}}
\bar Q_\ell(\tau)\bar r_\ell(\tau)\ge M_r^{\IDLE}(\bar{\boldsymbol Q}(\tau)).
\end{equation}
\end{lemma}

\begin{proof}
Write $b=b_n^{\mathrm{DL}}$ for the incoming downlink.
On an active backhaul, by the queue-eligibility constraint, both the queue and rate are at least $R_b^-$, so
\[
\min\{Q_b(t),R_b(m_n(\vs(t),t),t)\}s_b(t)
\ge R_b^-s_b(t).
\]
This proves \eqref{eq:FROM-gated-departure}.

Remove child backhauls violating the queue-eligibility constraints and empty access links from a maximizer of $W_n$ over $\widehat{\mathcal S}_n^{\theta_n(t)}(t)$, obtaining $\widetilde{\vs}_n$.
As the eliminated backhaul link's queue length and the rate are at most $\lceil R_{b_c^{\mathrm{DL}}}^-\rceil$, and $R_{b_c^{\mathrm{DL}}}^-$, respectively,
\begin{align*}
W_n(\widetilde{\vs}_n,t)
&\ge\max_{\vs_n\in\widehat{\mathcal S}_n^{\theta_n(t)}(t)}
W_n(\vs_n,t)-\sum_{c\in\mathcal C_n}
\lceil R_{b_c^{\mathrm{DL}}}^-\rceil R_{b_c^{\mathrm{DL}}}^-.
\end{align*}
The incoming term in FROM's objective lies between zero and $s_b(t)Q_b(t)R_b^+$.
Optimality in \eqref{eq:FROM-reduced-overhead-mw} therefore gives
\[
W_n(\vs_n^{\mathrm{FROM}}(t),t)
\ge W_n(\widetilde{\vs}_n,t)-s_b(t)Q_b(t)R_b^+.
\]
For an active backhaul, \eqref{eq:FROM-gated-departure} ensures that the actual departure is at least the minimum rate used in the objective.
For an active access link, if the queue contains less data than the link can transmit, the unused transmission capacity reduces its weight, $Q_\ell(t) R_\ell(t)$ by at most
\begin{align}
    &Q_\ell(t)R_\ell (t) - Q_\ell(t)\min\{Q_\ell(t), R_\ell(t)\} \nonumber\\
    &=Q_\ell(t)(R_\ell(t)-Q_\ell(t))^+\le R_{\max}^2/4,
\end{align}
where $R_{\max}$ is the maximum rate of all downlink rates.
Combining these bounds yields
\begin{align}
&\sum_{\ell\in\mathcal L_n^{\mathrm{TX}}}
Q_\ell(t)\sum_{f\in\mathcal F_\ell}d_n^f(t)
\nonumber\\
&\quad\ge\max_{\vs_n\in\widehat{\mathcal S}_n^{\theta_n(t)}(t)}
W_n(\vs_n,t)-s_b(t)Q_b(t)R_b^+ - O(1),
\label{eq:FROM-finite-weight-loss}
\end{align}
where the nonnegative $O(1)$ term is bounded by
\[
\sum_{c\in\mathcal C_n}
\lceil R_{b_c^{\mathrm{DL}}}^-\rceil R_{b_c^{\mathrm{DL}}}^-
+\frac{R_{\max}^2}{4}
|\mathcal L_n^{\mathrm{TX}}\cap\mathcal L_{\mathrm{AC}}|.
\]
Because the feasible set of $\theta = \IDLE$ contains all of the actions of $\theta = T$, $M_n^\IDLE(\boldsymbol Q)\ge M_n^T(\boldsymbol Q)$.

Under Assumption~\ref{assump:reduced-overhead-independence}, the parent's decision is independent of $n$'s other current SNRs conditional on the queues and incoming-backhaul availability.
Since $s_b(t)=1$ implies $R_b > 0$,
\begin{align}
&\mathbb E\Bigl[
\max_{\vs_n\in\widehat{\mathcal S}_n^{\IDLE}(t)}W_n(\vs_n,t)
-\max_{\vs_n\in\widehat{\mathcal S}_n^T(t)}W_n(\vs_n,t)\Bigm|\boldsymbol Q(t),s_b(t)=1\Bigr]
\nonumber\\
&\quad=M_n^\IDLE(\boldsymbol Q(t))-M_n^T(\boldsymbol Q(t)).
\label{eq:FROM-selected-channel-distribution}
\end{align}
Using $\theta_n(t)=T$ when $s_b(t)=1$ gives
\begin{align}
&\mathbb E\!\left[
\max_{\vs_n\in\widehat{\mathcal S}_n^{\theta_n(t)}(t)}
W_n(\vs_n,t)
\,\middle|\,\boldsymbol Q(t)\right]
\nonumber\\
&\quad=M_n^{\IDLE}(\boldsymbol Q(t))
-\mathbb E[s_b(t)\mid\boldsymbol Q(t)]\cdot[
M_n^\IDLE(\boldsymbol Q(t))-M_n^T(\boldsymbol Q(t))].
\label{eq:FROM-conditional-maximum}
\end{align}

The preceding scheduling bound involves conditional mean departures and activations, whereas the fluid limits in \eqref{eq:FROM-fluid-departure-limit} and \eqref{eq:FROM-fluid-activation-limit} are unconditional.
We will show that conditional expectation does not change the fluid limits.

Let $\mathcal H_t^x$ denote the history consisting of the queue lengths and all arrivals, channel states, scheduling decisions, and departures before slot $t$.
This history includes the current queues $\boldsymbol Q^x(t)$ but does not include the channel states in slot $t$.
For either $Y^x(t)=\sum_{f\in\mathcal F_\ell}d_n^{f,x}(t)$ or $Y^x(t)=s_b^x(t)$, since FROM determines its schedule using the current queues and channel states, and the channel states are independent across slots, we have
\[
\mathbb E[Y^x(t)\mid\mathcal H_t^x]
=
\mathbb E[Y^x(t)\mid\boldsymbol Q^x(t)].
\]
Consequently,
\[
M^x(k)\eqdef
\sum_{t=0}^{k-1}
\Bigl(
Y^x(t)
-\mathbb E[Y^x(t)\mid\boldsymbol Q^x(t)]
\Bigr)
\]
is a martingale with respect to $(\mathcal H_k^x)_{k\ge0}$.
Since $d_n^{f,x}$ and $s^x_b$ are bounded, there is a constant $C<\infty$ such that
\[
|M^x(t+1)-M^x(t)|\le C.
\]
For fixed $T>0$ and $\eta>0$, the Azuma--Hoeffding inequality gives, for sufficiently large $x$,
\begin{align*}
&\Pr\!\left\{
\frac1x\max_{0\le t\le\lfloor xT\rfloor}
|M^x(t)|\ge\eta
\right\}\\
&\quad\le
2\exp\!\left(
-\frac{\eta^2x^2}{2C^2\lfloor xT\rfloor}
\right)\\
&\quad\le
2\exp\!\left(
-\frac{\eta^2x}{2C^2T}
\right).
\end{align*}
Since the initial queue length $x_k$ diverges to infinity, we can select a subsequence from $x_k \geq k$.
Relabel this subsequence as $x_k$ and obtain $x_k \geq k$.
Then, the sum of probabilities is finite.
\[
\sum_{k=1}^{\infty}
2\exp\!\left(-\frac{\eta^2x_k}{2C^2T}\right)
\le
2\sum_{k=1}^{\infty}
\exp\!\left(-\frac{\eta^2k}{2C^2T}\right)
<\infty.
\]
The Borel--Cantelli lemma therefore gives the following convergence almost surely, which also holds along the fluid subsequence:
\[
\frac1{x_{k}}
\max_{0\le m\le\lfloor x_{k}T\rfloor}
|M^{x_{k}}(m)|
\longrightarrow0.
\]

For $\tau+h\le T$, set $x=x_{k_j}$ and sum over $t=\lfloor x\tau\rfloor,\ldots,\lfloor x(\tau+h)\rfloor-1$.
By the definition of $M^x$,
\begin{align*}
&\left|
\frac1x\sum_t
\mathbb E[Y^x(t)\mid\boldsymbol Q^x(t)]
-\frac1x\sum_tY^x(t)
\right|\\
&\quad=
\frac{
|M^x(\lfloor x(\tau+h)\rfloor)
-M^x(\lfloor x\tau\rfloor)|
}{x}\\
&\quad\le
\frac2x\max_{0\le m\le\lfloor xT\rfloor}|M^x(m)|
\longrightarrow0.
\end{align*}
Thus, using \eqref{eq:FROM-fluid-departure-limit} and \eqref{eq:FROM-fluid-activation-limit}, we obtain
\begin{align*}
&\frac1x\sum_t\mathbb E\!\left[
\sum_{f\in\mathcal F_\ell}d_n^{f,x}(t)
\,\middle|\,\boldsymbol Q^x(t)\right]\longrightarrow
\int_\tau^{\tau+h}\bar r_\ell(u)\,du,\\
&\frac1x\sum_t
\mathbb E[s_b^x(t)\mid\boldsymbol Q^x(t)]\longrightarrow
\int_\tau^{\tau+h}\bar s_b(u)\,du.
\end{align*}

From the definition of $M_n^\theta$, and for $\theta\in\{\IDLE,T\}$ and $\boldsymbol Q,\boldsymbol Q'\ge0$,
\[
|M_n^\theta(\boldsymbol Q)-M_n^\theta(\boldsymbol Q')|
\le R_{\max}\!
\sum_{\ell\in\mathcal L_n^{\mathrm{TX}}}
|Q_\ell-Q_\ell'|,
\]
Taking the conditional expectation of \eqref{eq:FROM-finite-weight-loss}, substituting \eqref{eq:FROM-conditional-maximum}, and summing after division by $x^2$ therefore yields
\begin{align*}
&\int_\tau^{\tau+h}
\sum_{\ell\in\mathcal L_n^{\mathrm{TX}}}
\bar Q_\ell(u)\bar r_\ell(u)\,du \ge\int_\tau^{\tau+h}
M_n^{\IDLE}(\bar{\boldsymbol Q}(u))\,du\\
&-\int_\tau^{\tau+h}\bar s_b(u)
\Bigl[M_n^\IDLE(\bar{\boldsymbol Q}(u))
-M_n^T(\bar{\boldsymbol Q}(u))+\bar Q_b(u)R_b^+\Bigr]\,du.
\end{align*}
The constant term independent of queue lengths vanishes as $j\to\infty$.
Dividing by $h$ and letting $h\rightarrow0$ proves \eqref{eq:FROM-fluid-weight-before-relaxation}.
\end{proof}

\begin{lemma}
\label{lem:FROM-reference-service}
For the mean service $\overline c_\ell$ chosen in \eqref{eq:reduced-overhead-policy-initial-slack}, the root satisfies
\begin{equation}
\label{eq:FROM-root-reference-bound}
M_r^{\IDLE}(\bar{\boldsymbol Q}(\tau))\ge
\beta\sum_{\ell\in\mathcal L_r^{\mathrm{TX}}}
\bar Q_\ell(\tau)\overline c_\ell.
\end{equation}
Every non-root IAB satisfies
\begin{align}
&M_n^{\IDLE}(\bar{\boldsymbol Q}(\tau))
-\rho_n\left(M_n^\IDLE(\bar{\boldsymbol Q}(\tau))-M_n^T(\bar{\boldsymbol Q}(\tau))\right)
\nonumber\\
&\qquad\ge
\beta\sum_{\ell\in\mathcal L_n^{\mathrm{TX}}}
\bar Q_\ell(\tau)\overline c_\ell.
\label{eq:FROM-reference-mode-bound}
\end{align}
\end{lemma}

\begin{proof}
In this lemma's proof, all link sums below are over $n$'s outgoing downlinks.
For each $d_j$, let $\vs_n$ be its schedule at $n$ with outaged links turned off, and let $\theta$ be the corresponding parent-backhaul mode.
By the definition of $\beta$, each child backhaul satisfies
\[
R_{b_c^{\mathrm{DL}}}^-s_{b_c^{\mathrm{DL}}}^{d_j}(t)
\mathbf1\{c_{b_c^{\mathrm{DL}}}^{d_j}(t)>0\}
\ge\beta c_{b_c^{\mathrm{DL}}}^{d_j}(t).
\]
Therefore,
\begin{align} \label{ineq:FROM-beta-W}
\beta\sum_\ell\bar Q_\ell(\tau)c_\ell^{d_j}(t)
&\le W_n(\vs_n,t;\bar{\boldsymbol Q}(\tau))\\
&\le
\max_{\widetilde{\vs}_n\in\widehat{\mathcal S}_n^\theta(t)}
W_n(\widetilde{\vs}_n,t;\bar{\boldsymbol Q}(\tau)).
\end{align}
At the root, $\theta_r=\IDLE$ and taking expectations and summing with $\alpha_j$ gives
\[
M_r^{\IDLE}(\bar{\boldsymbol Q}(\tau))
\ge\beta\sum_\ell\bar Q_\ell(\tau)
\sum_{j=1}^K\alpha_j\mathbb E[c_\ell^{d_j}(t)]
=\beta\sum_\ell\bar Q_\ell(\tau)\overline c_\ell.
\]
For $n\ne r$, similarly take expectations in \eqref{ineq:FROM-beta-W} and consider the cases $\theta = T$ and $\theta = \IDLE$ separately.
As the parent's decision is independent of its children's information,
\begin{align}
&\beta\mathbb E\!\left[
\sum_\ell\bar Q_\ell(\tau)c_\ell^{d_j}(t)\right]
\nonumber\\
&\le
\mathbb E\!\left[
\mathbf1\{\theta=\IDLE\}
\max_{\widetilde{\vs}_n\in\widehat{\mathcal S}_n^{\IDLE}(t)}
W_n(\widetilde{\vs}_n,t;\bar{\boldsymbol Q}(\tau))
\right]
\nonumber\\
&\quad+\Pr\{\theta=T\}M_n^T(\bar{\boldsymbol Q}(\tau))
\nonumber\\
&=
M_n^{\IDLE}(\bar{\boldsymbol Q}(\tau))
-\Pr\{\theta=T\}M_n^\IDLE(\bar{\boldsymbol Q}(\tau))
\nonumber\\
&\quad+\Pr\{\theta=T\}M_n^T(\bar{\boldsymbol Q}(\tau))
\nonumber\\
&=
M_n^{\IDLE}(\bar{\boldsymbol Q}(\tau))
-\mathbb E\!\left[
s_{b_n^{\mathrm{DL}}}^{d_j}(t)
\mathbf1\{R_{b_n^{\mathrm{DL}}}(0,t)>0\}
\right]
\nonumber\\
&\quad\cdot\bigl[
M_n^\IDLE(\bar{\boldsymbol Q}(\tau))
-M_n^T(\bar{\boldsymbol Q}(\tau))
\bigr].
\label{eq:FROM-rule-mode-bound}
\end{align}
where the second last line comes from \eqref{eq:FROM-conditional-maximum}.
\end{proof}

\begin{lemma}
\label{lem:FROM-ancestor-service}
Suppose all outgoing fluid queues at depths smaller than $i$ remain zero for $\tau\ge t_i$.
For a non-root IAB $n$ at depth $i$, $\bar Q_{b_n^{\mathrm{DL}}}(\tau)=0$, and at every regular time $\tau>t_i$,
\begin{equation}
\label{eq:FROM-ancestor-conservation}
\bar r_{b_n^{\mathrm{DL}}}(\tau)=\lambda_{b_n^{\mathrm{DL}}},
\qquad \bar a_\ell(\tau)=\lambda_\ell,
\end{equation}
for $\ell\in\mathcal L_n^{\mathrm{TX}}$.
Moreover, $\bar s_{b_n^{\mathrm{DL}}}(\tau)\le\rho_n$.
\end{lemma}

\begin{proof}
The fluid limit of parent's queue is zero by the hypothesis.
For an ancestor $a$ and flow $f$ using its outgoing link $\ell$, $0\le q_a^{f,x}(\lfloor xu\rfloor)/x\le Q_\ell^x(\lfloor xu\rfloor)/x\to0$ for $u\ge t_i$.
Hence, the queue equation gives, for $t_i\le u<v$,
\begin{align}
&\frac1x\sum_{t=\lfloor xu\rfloor}^{\lfloor xv\rfloor-1}
\bigl(a_a^{f,x}(t)-d_a^{f,x}(t)\bigr)
\nonumber\\
&\quad=\frac{q_a^{f,x}(\lfloor xv\rfloor)-q_a^{f,x}(\lfloor xu\rfloor)}x
\longrightarrow0.
\label{eq:FROM-per-flow-conservation}
\end{align}
From $u$ to $v$, $\lambda_f(v-u)$ packets arrive at the root.
Applying \eqref{eq:FROM-per-flow-conservation} and $a_a^f(t)=d_{\mathrm{pr}(a)}^f(t)$ along the route gives the same quantity for each flow delivered to $n$.
Summing over the flows on each link and differentiating gives \eqref{eq:FROM-ancestor-conservation}.
The fluid limit of \eqref{eq:FROM-gated-departure}, together with \eqref{eq:FROM-reference-slack} and \eqref{eq:FROM-reference-upper-bound}, gives
\begin{align}
\bar s_{b_n^{\mathrm{DL}}}(\tau)
&\le\frac{\bar r_{b_n^{\mathrm{DL}}}(\tau)}{R_{b_n^{\mathrm{DL}}}^-}
=\frac{\lambda_{b_n^{\mathrm{DL}}}}{R_{b_n^{\mathrm{DL}}}^-}
\nonumber\\
&\le\frac{\beta\overline c_{b_n^{\mathrm{DL}}}}{R_{b_n^{\mathrm{DL}}}^-}
\le\frac{\overline c_{b_n^{\mathrm{DL}}}}{R_{b_n^{\mathrm{DL}}}^+}
\le\rho_n.
\label{eq:FROM-mode-occupation}
\end{align}
\end{proof}

\begin{proof}[Proof of Theorem~\ref{thm:FROM-stability}]
We prove by induction on node depth that all fluid queues become zero in finite time and remain zero.
Define
\begin{equation}
\label{eq:FROM-local-lyapunov}
V_n(\tau)\eqdef
\sum_{\ell\in\mathcal L_n^{\mathrm{TX}}}
\bar Q_\ell(\tau)^2.
\end{equation}

\emph{Base case.}
At the root, $\bar a_\ell(\tau)=\lambda_\ell$.
Lemmas~\ref{lem:FROM-fluid-service} and~\ref{lem:FROM-reference-service} give $\sum_\ell\bar Q_\ell\bar r_\ell\ge\beta\sum_\ell\bar Q_\ell\overline c_\ell$, with sums over the root's outgoing downlinks.
Using \eqref{eq:FROM-fluid-balance} and \eqref{eq:FROM-reference-slack},
\begin{align*}
\sum_\ell\bar Q_\ell(\tau)\bar r_\ell(\tau)
&\ge
M_r^{\IDLE}(\bar{\boldsymbol Q}(\tau))\ge
\beta\sum_\ell\bar Q_\ell(\tau)\overline c_\ell,
\end{align*}
where the sums are over the root's outgoing downlinks.
Using \eqref{eq:FROM-fluid-balance} and \eqref{eq:FROM-reference-slack},
\begin{align}
\tfrac12\dot V_r(\tau)
&\le\sum_\ell\bar Q_\ell(\tau)(\lambda_\ell-\beta\overline c_\ell)
\nonumber\\
&\le-\delta\sum_\ell\bar Q_\ell(\tau)
\le-\delta\sqrt{V_r(\tau)}.
\label{eq:FROM-root-drift}
\end{align}
The $\sqrt{V_r}$ has derivative at most $-\delta$ almost everywhere where positive.
Thus, by solving the differential inequalities, we can get
\begin{equation}
\label{eq:FROM-root-draining}
\sqrt{V_r(\tau)}\le[\sqrt{V_r(0)}-\delta\tau]^+.
\end{equation}
The root's fluid queues are therefore zero for all $\tau\ge\sqrt{V_r(0)}/\delta$.

\emph{Induction step.}
Suppose all outgoing queues at depths smaller than $i$ remain zero after $t_i$, and let $n$ have depth $i$.
Lemma~\ref{lem:FROM-ancestor-service} gives $\bar Q_{b_n^{\mathrm{DL}}}=0$, $\bar a_\ell=\lambda_\ell$, and $\bar s_{b_n^{\mathrm{DL}}}\le\rho_n$.
Since $M_n^\IDLE(\bar{\boldsymbol Q}(\tau))\ge M_n^T(\bar{\boldsymbol Q}(\tau))$, replace $\bar s_{b_n^{\mathrm{DL}}}$ by its upper bound $\rho_n$ in Lemma~\ref{lem:FROM-fluid-service} and apply Lemma~\ref{lem:FROM-reference-service} to obtain
\begin{align}
\sum_\ell\bar Q_\ell(\tau)\bar r_\ell(\tau)
&\ge M_n^{\IDLE}(\bar{\boldsymbol Q}(\tau))-\rho_n\Bigl[M_n^\IDLE(\bar{\boldsymbol Q}(\tau))
-M_n^T(\bar{\boldsymbol Q}(\tau))\Bigr]
\nonumber\\
&\ge\beta\sum_\ell\bar Q_\ell(\tau)\overline c_\ell,
\label{eq:FROM-local-dominance}
\end{align}
where the sums are over $n$'s outgoing downlinks.
The same calculation as \eqref{eq:FROM-root-drift} yields $\dot V_n/2\le-\delta\sqrt{V_n}$ and thus
\begin{equation}
\label{eq:FROM-node-draining}
\sqrt{V_n(\tau)}\le
[\sqrt{V_n(t_i)}-\delta(\tau-t_i)]^+,
\quad\tau\ge t_i.
\end{equation}

Using the common backlog bound \eqref{eq:FROM-total-fluid-bound}, set $t_0=0$ and
\begin{equation}
\label{eq:FROM-uniform-drain-time}
t_{i+1}=t_i+\frac{1+t_i\sum_{f\in\mathcal F_{\mathrm{DL}}}\lambda_f}{\delta}.
\end{equation}
Since $\sqrt{V_n(t_i)}\le\sum_\ell\bar Q_\ell(t_i)$, all queues transmitted at depth $i$ remain zero from $t_{i+1}$ onward.
The finite tree depth gives a common draining time for all normalized fluid limits.
Hence, by \cite{dai1995stability}, FROM can stabilize the network.
\end{proof}

\bibliographystyle{bibtex/IEEEtran}
\bibliography{bibtex/IEEEabrv,refs}

\end{document}